\documentclass[10pt,journal,cspaper,compsoc]{IEEEtran}

\ifCLASSOPTIONcompsoc
\usepackage[nocompress]{cite}
\else
\usepackage{cite}
\fi

\usepackage{graphicx}
\usepackage{subfigure}
\usepackage{epsfig}
\usepackage{amsmath}
\usepackage{amssymb}
\usepackage{cases}
\usepackage{hhline}
\usepackage{caption}
\usepackage{algorithm}
\usepackage{algorithmic}
\usepackage[T1]{fontenc}
\usepackage{color}

\begin{document}

\title{Deep Reinforcement Learning for Privacy-Preserving Task Offloading in Integrated Satellite-Terrestrial Networks}

\author{Wenjun Lan, Kongyang Chen, Yikai Li, Jiannong Cao,~\IEEEmembership{Fellow,~IEEE}, Yuvraj Sahni
\IEEEcompsocitemizethanks{
\IEEEcompsocthanksitem{W. Lan, K. Chen, and Y. Li are with Institute of Artificial Intelligence and Blockchain, Guangzhou University, China. K. Chen is also with Pazhou Lab, Guangzhou, China. E-mail: kychen@gzhu.edu.cn. (Corresponding author: K. Chen)}
\IEEEcompsocthanksitem{J. Cao is with Department of Computing, The Hong Kong Polytechnic University, Hong Kong, China.} 
\IEEEcompsocthanksitem{Y. Sahni is with Department of Building Environment and Energy Engineering, The Hong Kong Polytechnic University, Hong Kong, China.} 
}}

\IEEEtitleabstractindextext{
\begin{abstract}
Satellite communication networks have attracted widespread attention for seamless network coverage and collaborative computing. In satellite-terrestrial networks, ground users can offload computing tasks to visible satellites that with strong computational capabilities. Existing solutions on satellite-assisted task computing generally focused on system performance optimization such as task completion time and energy consumption. However, due to the high-speed mobility pattern and unreliable communication channels, existing methods still suffer from serious privacy leakages. In this paper, we present an integrated satellite-terrestrial network to enable satellite-assisted task offloading under dynamic mobility nature. We also propose a privacy-preserving task offloading scheme to bridge the gap between offloading performance and privacy leakage. In particular, we balance two offloading privacy, called the usage pattern privacy and the location privacy, with different offloading targets (e.g., completion time, energy consumption, and communication reliability). Finally, we formulate it into a joint optimization problem, and introduce a deep reinforcement learning-based privacy-preserving algorithm for an optimal offloading policy. Experimental results show that our proposed algorithm outperforms other benchmark algorithms in terms of completion time, energy consumption, privacy-preserving level, and communication reliability. We hope this work could provide improved solutions for privacy-persevering task offloading in satellite-assisted edge computing.
\end{abstract}

}
	
\maketitle
\IEEEdisplaynontitleabstractindextext
\IEEEpeerreviewmaketitle

\section{Introduction}\label{sec:introduction}
In recent years, satellite technology has made significant improvements across commercial, civil, and military fields, primarily driven by the development of space communication networks. Satellites are also serving as one of the crucial components in 6G networks \cite{you2021towards, giordani2020non, chen2020system, zhu2021integrated}, especially the Low Earth orbit (LEO) satellites with small sizes, low costs, and easy deployments. Both academia and industry are interested in the satellite constellations deployment for global and seamless network coverage \cite{chaudhry2021laser,9686591}. 

As the continuous development and enrichment of satellite services, satellite-terrestrial networks enable more and more opportunities for collaborative computing among the satellites and various types of ground equipment. For example, ultra-small satellites (e.g., remote sensing satellites) usually transmit their data samples to terrestrial data centers for intelligent data processing. In recent decades, many satellites are also equipped with high-performance computation resources, generating more satellite-terrestrial cooperation scenarios. 

Thus, the computational capabilities of available satellites have been gradually exploited to facilitate resource-constrained ground user equipment. For this purpose, recent studies have integrated edge computing into satellite networks. Edge computing is a promising approach that leverages the computational capabilities of network edge nodes for efficient data processing \cite{mao2017survey, 9827596, 9591418, 9382907}. By integrating edge computing into satellite-terrestrial networks, ground users can offload computation tasks to their nearby visible satellites to enable more efficient service guarantees for latency-sensitive and compute-intensive applications. 
However, task offloading policies are usually constrained by various factors such as device capabilities, network conditions, server capacities, and user requirements (e.g., completion time, energy consumption, privacy). 
Also, in integrated satellite-terrestrial networks, the altitude limitation of satellites generates increased transmission delays for terrestrial users, presenting challenges in fulfilling real-time communication requirements. 

Previous solutions on satellite-assisted edge computing have predominantly focused on optimizing system performance, such as completion time and energy consumption, during task offloading. However, two critical issues have often been overlooked in existing solutions. \textit{The first issue is the potential exposure of the user's private information}. Offloading tasks to an insecure satellite may expose the user's location privacy and usage pattern privacy. For location privacy, insecure satellites can infer communication channel states to obtain the user's location, as channel states are highly correlated with the distances and locations among the UE and satellites. For usage pattern privacy, insecure satellites might extract statistical information as a unique identifier or fingerprint of the user, directly from its task offloading history. \textit{The second issue is the high mobility and the limited coverage time of each satellite}. Each satellite can only provide services to the user at specific time periods or locations. If a satellite becomes invisible, it should migrate its computation results to the nearest satellite within the visible range, which transmits these results back to the user to ensure a reliable backhaul transmission.

To tackle these issues in the satellite-assisted edge computing architecture, in this paper, we propose an integrated satellite-terrestrial network to optimize the offloading cost, communication reliability and user privacy leakage, subject to certain constraints such as satellite mobility and coverage time. We further formulate this problem with a Markov Decision Process (MDP) model. We also propose a PPO-based deep reinforcement learning algorithm to achieve an optimal task offloading policy, which minimizes the total time and energy consumption required for the task computation, in accordance with the requirements of a pre-specified privacy level and a reliability level. 

To the best of our knowledge, we are the first work to consider the high mobility of satellite movement as well as the privacy-preserving requirement for satellite-assisted task offloading. The main contributions of this paper can be summarized as follows:

\begin{enumerate}
\item We propose an integrated satellite-terrestrial network, which leverages the computational capabilities of LEO satellites to provide efficient computing services for terrestrial users. Compared with existing solutions, we also consider the dynamic nature of satellite mobility, and present a backhaul migration mechanism to support offloading results transmission for invisible satellites.
	
\item We present a privacy-preserving task offloading scheme in the satellite-terrestrial network to bridge the gap between offloading performance and privacy leakage in this area. In particular, we balance two offloading privacy metrics (i.e., the usage pattern privacy and the location privacy) with traditional satellite offloading targets (e.g., computation time, energy consumption, and communication reliability), and determine a joint optimization framework. 
		
\item We propose a  deep reinforcement learning-based privacy-preserving task offloading algorithm for an optimal offloading policy. Experimental results show that our proposed algorithm outperforms other benchmark algorithms, producing an excellent balance among task completion time, energy consumption, privacy-persevering level, and communication reliability. 
\end{enumerate}

The remainder of this paper is organized as follows. Section \ref{sec:Related Work} gives an overview of the related work. Section \ref{sec:System Model} introduces our satellite-terrestrial network for privacy-preserving task offloading. Section \ref{sec:Privacy-preserving Task Offloading Algorithm} presents a deep reinforcement learning algorithm to achieve an optimal task offloading policy. Section \ref{sec:Performance Evaluation} shows the experimental results. Finally, Section \ref{sec:conclusion} concludes this paper.

\section{Related Work}\label{sec:Related Work}

\subsection{Satellite-Assisted Task Offloading}
Task offloading is an efficient computation approach for many resource-constrained devices \cite{huang2020result,9835126}. 
Inspired by MEC technology, LEO satellite networks will also leverage their abundant computational resources to enable the provisioning of computing services to UEs. 
To exploit the computation abilities of LEO satellites, \cite{song2021energy} introduced a satellite-terrestrial-based edge computing solution that offloads tasks from IoT devices to nearby satellites for fast computation.  
Tang et al.\cite{tang2021computation} proposed a hybrid cloud-edge assisted architecture where LEO satellite networks offer heterogeneous computing resources to minimize the UE's energy consumption.
Wang et al.\cite{wang2020joint} addressed a joint computation offloading and resource allocation scheme in an LEO satellite-based edge computing system, producing a mixed-integer nonlinear programming problem with a game theory and Lagrange multiplier operator.
Cheng et al. \cite{cheng2019space} formulated the task offloading decision in a satellite-UAV-served IoT network as an MDP model under network dynamics, determining an optimal computation offloading policy with Deep Reinforcement Learning (DRL).
In \cite{xu2020deep}, a space-air-ground-sea integrated network, constituted of UEs, unmanned aerial vehicles (UAVs), and LEO satellites,  was investigated to facilitate hybrid computing services in maritime IoT applications. Their computation and communication resources are also jointly allocated by DRL.
Wang et al. \cite{wang2018computation} introduced a novel satellite-terrestrial network architecture to enable double edge computing, where MEC servers are deployed on terrestrial base stations as well as the satellite network, to tackle the challenge of limited computing resources in edge servers in remote areas. In this work, a minimum cost matching algorithm is employed to optimize the global energy consumption and the average latency on edge servers.
Zhang et al.\cite{zhang2019satellite} deployed a dynamic Network Functions Virtualization (NFV) system to integrate computing resources within the coverage area of LEO Satellites. 
They also proposed a collaborative computation offloading method for Satellite-MEC scenarios to reduce the user perception delay and energy consumption.
Furthermore, Qiu et al. \cite{qiu2019deep} proposed a software-defined satellite-ground network to dynamically manage their cache and computation resources, solving the joint resource allocation optimization problem with a deep Q-learning algorithm.

In summary, these satellite-based edge computing systems generally focus on task offloading schemes in different application domains. 
However, the satellite mobility problem is not considered in existing solutions. 
Also, little attention has been paid to the privacy leakage problem during the task offloading process. 
To the best of our knowledge, we are the first for joint satellite mobility patterns and offloading privacy in the satellite-assisted edge computing area.

\subsection{Privacy-Preserving Task Offloading}

Privacy-preserving task offloading has become an emerging research hotspot in recent years. 
It has been confirmed that the task offloading process will lead to potential privacy leakages, where an adversary can infer user locations and usage patterns by monitoring offloading decisions. 
During the task offloading process, there are several privacy metrics to measure the privacy-preserving level. 
The first one is so-called \textit{privacy entropy}, extending from traditional information entropy \cite{xu2019privacy}. 
For example,  privacy entropy is considered as an optimization objective in \cite{xu2019time}, and also referred to quantify the probability of being attacked in offloading tasks \cite{9293144}. 
There are also many other privacy metrics.
In \cite{xu2019privacy}, it proposed a heuristic privacy metric to jointly quantify the location privacy and the usage pattern privacy of mobile users, where the task offloading process is modeled as a constrained Markov decision process (CMDP) to minimize the energy consumption and the latency under a pre-specified privacy level.
In \cite{min2018learning}, it further studied the usage pattern privacy and the location privacy in healthcare IoT scenarios, and proposed a reinforcement learning algorithm to jointly optimize the privacy level, computational latency, and energy consumption.
In\cite{li2020privacy}, it further extended the idea of \cite{min2018learning} by removing the prior knowledge of any system-level information.
In \cite{dong2020privacy}, it measured privacy leakages by comparing the number of locally computing tasks and the number of offloading tasks.
 Other privacy metrics are also considered, such as task sensitivity, location privacy loss \cite{he2017location, wang2018not}, and differential privacy \cite{xu2018distilling}. 
 For example, in \cite{pang2022towards}, it provided user location privacy protection by interfering with the distance between the user and the edge server with differential privacy.

In summary, these solutions mainly focus on privacy-preserving task offloading in traditional edge computing scenarios. However, these solutions are not well-suited to the satellite-assisted edge computing systems, due to complicated environmental constraints and high mobility patterns in satellite systems.

\section{System Model}\label{sec:System Model}
In this section, we briefly introduce the integrated satellite-terrestrial network for users' task offloading. We also propose several specific models for each component including satellite coverage, user channel, time delay, energy consumption, transmission reliability, and offloading privacy. Finally, we formulate our task offloading problem.

\subsection{Integrated Satellite-Terrestrial Network-Assisted Edge Computing Architecture}

\begin{figure}[!t]
\centering
\includegraphics[width=0.8\columnwidth]{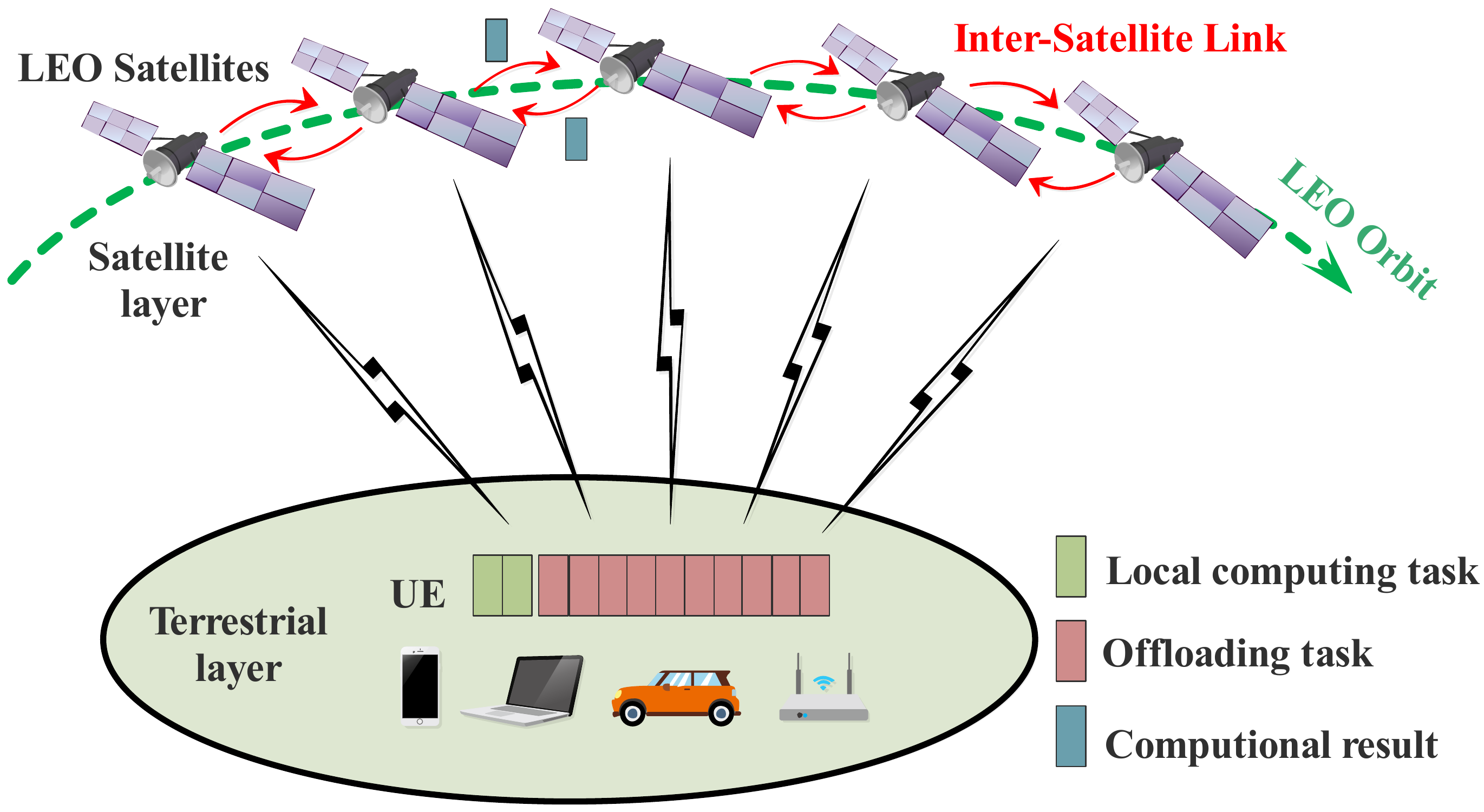}
\caption{Integrated satellite-terrestrial network.}
\label{fig:Satellite-Terrestrial Network Edge Computing Architecture}
\end{figure}

In this paper, we propose an integrated satellite-terrestrial network-assisted edge computing architecture to support an energy-efficient, low latency, and privacy-preserving task offloading scheme, as shown in Figure \ref{fig:Satellite-Terrestrial Network Edge Computing Architecture}. The architecture contains two main components: a satellite layer consisting of LEO satellites that are equipped with MEC servers, and a terrestrial layer where a UE generates the computational tasks. We consider a scenario with a single UE and multiple LEO satellites, where the UE can transmit directly to the LEO satellites with wireless communication signals. Each satellite is flying at high speed over the Earth, equipped with a MEC server to provide distributed computing capabilities. Computational results can be migrated between these satellites via inter-satellite links (ISL).

The communication time between UE and LEO satellites is limited by the coverage time of LEO satellites, so LEO satellites can only provide computing services for the UE under certain circumstances. Therefore, we employ a partial offloading scheme, which means computational tasks are divided into two parts: one executed on the UE and the other executed on the satellites. 
Computational tasks can be reasonably allocated between the UE and satellites to achieve efficient offloading of computational tasks. Thus, the UE has two offloading policies to handle each computational task, i.e., the UE can keep the task locally for computation or offload the task to the satellites for computation. The task distribution in the UE offloading follows a time-sequential scheme, where each task is transmitted in order, utilizing the entire channel bandwidth, rather than simultaneously. Therefore, a reasonable task offloading order can optimize the task execution efficiency and reduce the total cost.

Assume that the UE has $N$ tasks, these tasks are indivisible, and the data size of task $i$ is denoted by $D_i$, where $i=1,2,\dots, N$. The UE also has a certain computational capacity, and the local computation speed is $\alpha$. Suppose there are $M$ satellites and the computation speed of the $j$th satellite is $\beta_{j}$, where $j=1,2,\dots,M$. 

We denote the task offloading decision by $\eta_{ij}$, where $\eta_{ij}=1$ if the UE offloads task $i$ to the satellite $j$, and $\eta_{ij}=0$ otherwise. Based on the offloading decision for task $i$, the flag $g_{i}=\Sigma_{j=1}^{M} \eta_{i j}$ is defined for whether task $i$ is offloaded to satellites. If $g_i=1$, it means task $i$ is offloaded; if $g_i=0$, it means task $i$ is for local computation. The other symbols used in the paper are listed in Table \ref{table1}. 

\begin{table}[!t] 
\centering
\caption{Summary of notations in our system.}
\footnotesize
%\small
%\normalsize
\begin{tabular}{p{1.4cm}p{7cm}} 
\hline 
notation &  description\\
\hline
$ N $ & {Total number of tasks} \\
$ M $ & {Total number of satellites} \\
$ i $ & {Task number} \\
$ j $ & {Satellite number} \\
$ D_i $ & {Data size of the task $i$} \\
$ \alpha $ & { Local computing speed} \\
$ \beta_{j} $ & {Computing speed of the satellite $j$} \\
$ \eta_{ij} $ & {Task offloading decision} \\
$ g_{i} $ & {Flag for offloading the task $i$ to satellites} \\
$ R $ & {The radius of the Earth} \\
$ H $ & {Orbital altitude of LEO satellite operation} \\
$ V $ & {Satellite movement speed} \\
$ \gamma  $ & {The angle between the line connecting the satellite and the geocenter and the clockwise direction of the reference line} \\
$ s_{j}(t) $ & {Distance between the UE and the satellite $j$ at time $t$ } \\
$h_{j}(t) $ & {  Channel gain between the UE and the satellite $j$ at time $t$ } \\
$ SNR_{j}(t) $ & {  SNR of the link between the UE and the satellite $j$ at time $t$   } \\
$	R_{j}(t) $ & {  Data transmission rate of the link between the UE and the satellite server $j$ at time $t$   } \\
$b_{j}(t) $ & {BER of transmitting the task to the satellite $j$ at time $t$} \\
$ t_{i,j}^{upload} $ & {Upload time to offload the task $i$ to the satellite $j$ at time $t$} \\
$ t_{i,j}^{comp} $ & {Computation time of the task $i$ on the satellite $j$} \\
$ t_{i}^{comp} $ & {Computation time of the task $i$ on the UE} \\
$ t_{local}^{comp} $ & {Computation time for all local tasks} \\
$ t^{upload,end} $ & {End time of the offloading transmission of the UE} \\
$ t_{local}^{comp,start} $ & {Start time of local computation by the UE} \\
$ t_{local}^{comp,end} $ & {End time of local computation by the UE} \\
$ M_{p} $ & {Task received by the satellite $p$} \\
$ t_{kp}^{comp,start} $ & {Computation start time of the task $M^k_p$ } \\
$ t_{kp}^{comp,end} $ & {Computation end time of the task $M^k_p$ } \\
$ t^{migrate}_{i} $ & {Time to migrate the computation result of task $i$} \\
$ t_{i}^{migrate,end} $ & {Migration end time of the computation result of the task $i$} \\
$ t_{i,j}^{download} $ & {Time to backhaul the computation result of the task $i$} \\
$ t_{i,j}^{download,end} $ & {End time of computation result to backhaul for the task $i$} \\
$ t_{i}^{end} $ & {Completion time of the task $i$} \\
$ T_{total} $ & {Total time to complete all tasks} \\
$ E^{comp} $ & {Energy consumption generated by the computation of the UE} \\
$ E^{tran} $ & {Energy consumption generated by the offloading transmission of the UE} \\
$ E $ & {Total energy consumption of the UE} \\
$ r_{i}^ {success} $ & {Probability of successful transmission of the task $i$} \\
$ r^{success} $ & {Probability of successful offloading of the UE} \\
$ r^{failure} $ & {Probability of failure offloading of the UE} \\
$ P_{u}(n) $ & {Usage pattern privacy-preserving rewards for the $n$-th offloading decision} \\
$ P_{l}(n) $ & {Location privacy-preserving rewards for the $n$-th offloading decision} \\
$ P_{total}(n) $ & {Total privacy-preserving level for the $n$-th offloading decision} \\
$ P_{total} $ & {Total privacy-preserving level of the UE} \\
$ C $ & {Total cost of the UE} \\
\hline
\end{tabular}
\label{table1}    
\end{table}

\subsection{Satellite Coverage Model}
We assume a uniform distribution of LEO satellites located in the same orbital plane. Each satellite's position dynamically varies as it orbits the Earth at a consistent angular speed $V$. %In contrast to MEO and GEO satellites, LEO satellites have lower orbital altitudes but higher velocities. 
According to \cite{tang2021computation,tong2022joint}, LEO satellites can only provide services to the UE at specific time periods or locations. Consequently, the UE cannot communicate with LEO satellites at all times and can only communicate when a specific relationship is satisfied. 

\begin{figure}[!t]
\centering
\includegraphics[width=0.75\columnwidth]{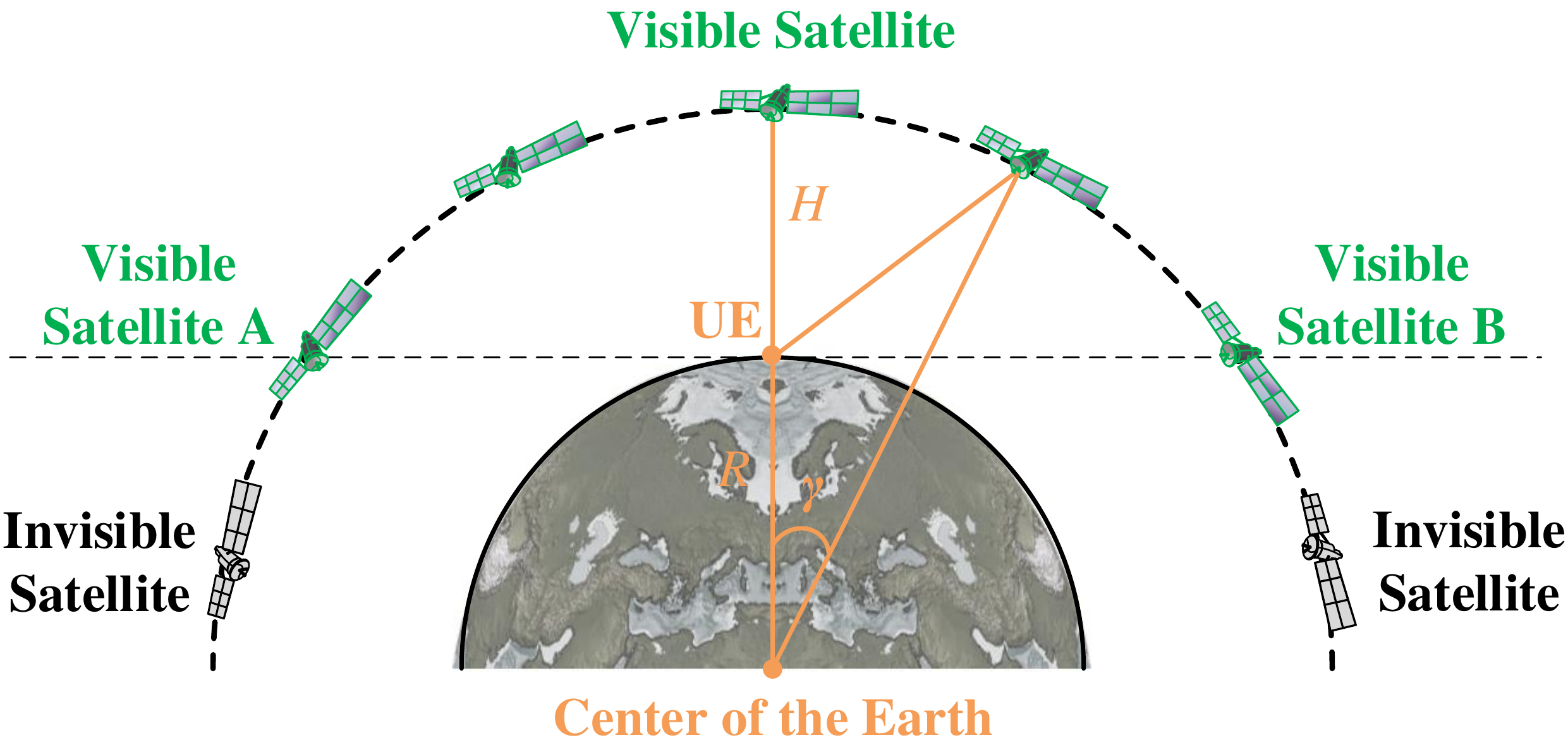}
\caption{Satellite coverage model.}
\label{fig:Satellite Coverage Model}
\end{figure}

Figure \ref{fig:Satellite Coverage Model} illustrates the spatial geometric relationship between LEO satellites and the UE. The reference line is defined as the line connecting the UE and the geocenter. The angle $\gamma$ represents the position of satellites, measured in the clockwise direction from the reference line. This angle is used to calculate the distance between the user and the satellite. The visible range of the satellite is depicted as the arc $\stackrel\frown{AB}$ in Figure \ref{fig:Satellite Coverage Model}. Only satellites in this range can receive tasks transmitted from the UE. 
If the satellite is invisible, it becomes necessary to migrate the computation result data to the nearest satellite within the visible range and transmit it back to ensure a reliable backhaul transmission.

Suppose that $\gamma$ represents the angle between the line connecting satellite $j$ and the geocenter, measured in the clockwise direction from the reference line at time $t$. In the triangle formed by the geocenter, the UE position, and the position of satellite $j$, the distance between the UE and satellite $j$ at time $t$ can be expressed using the cosine theorem as $s_{j}(t)=\sqrt{R^2+(R+H)^2-2R(R+H)\cos(\gamma)}$,
where $R$ is the radius of the Earth, and $H$ is the orbital altitude at which the LEO satellite operates.

\subsection{Satellite-Terrestrial Channel Model}\label{sec:Satellite-Terrestrial Channel Model}
Due to the continuous movement of satellites, the wireless transmission conditions also change constantly, bringing in significant challenges for satellite task offloading process. 
The transmission link utilizes the Ka-band for wireless communication, which ensures channel robustness in the presence of rainfall dynamics. Therefore, the channel conditions are primarily influenced by the communication distance. Since our paper focuses on a single UE, there is no interference among UEs. To facilitate analysis, the widely used free-space path loss model is adopted. This model neglects small-scale fast fading since the satellite is positioned at a high altitude, and line-of-sight (LoS) propagation dominates. Moreover, the Doppler effect resulting from satellite mobility is assumed to be perfectly compensated at the UE. %/satellite receiver (SR).

Therefore, assuming that the distance between the UE and the satellite $j$ is $s_{j}(t)$ at time $t$, the channel gain between the UE and the satellite $j$ is $h_{j}(t) = \frac {\beta _{o}}{{s_{j}(t)}^{2}},$
where $\beta _{o}$ denotes the channel gain at a reference distance of 1 m.

To ensure communication quality and reliability in a satellite-terrestrial system, it is assumed that each satellite utilizes distinct frequency bands during transmission to prevent signal interference. The Signal-to-Noise Ratio (SNR), denoting the ratio of signal strength to environmental noise, for the link between the UE and satellite $j$ at time $t$ is $SNR_{j}(t)=\frac {P^{tran}{h_{j}(t)}} {N_{0}},$
where $P^{tran}$ and $N_{0}$ denote the transmission power of the UE and noise power, respectively.

To determine the achievable data transmission rate for computational offloading between the UE and the satellite $j$ at time $t$, combined with the Shannon formula, this can be denoted as $R_{j}(t) = B{\log _{2}}\left ({{1 + SNR_{j}(t) }} \right),$
where $B$ denotes the channel bandwidth from the link between the UE and the satellite.

In addition, we can assess the channel state condition by evaluating the magnitude of the channel gain, denoted as $h(t)$. A channel gain exceeding a pre-specified threshold signifies a favorable channel state condition, while a channel gain below the threshold indicates a poor channel state condition. This method facilitates a more comprehensive understanding of the channel conditions in satellite communication systems. By considering the dynamic changes in the channel state during task offloading decisions, we can enhance the performance of task offloading significantly.

In our paper, the Bit Error Rate (BER) is utilized as a metric to quantify the probability of data corruption for a single bit in the transmission system at time $t$. A significant correlation exists between the theoretical BER and the SNR. Specifically, as the SNR increases, the BER decreases. We assume that the satellite communication system employs Binary Phase Shift Keying (BPSK) modulation. The BER for transmitting the task to the satellite server $j$ at time $t$ can be determined by $b_{j}(t)=\frac{erfc(\sqrt{SNR_{j}(t)} )}{2},$
where $erfc()$ is the complementary error function. In this way, we can estimate the BER during task transmission under varying SNR conditions. Accurate BER estimation enables a comprehensive evaluation of transmission quality in the satellite communication system at different time intervals, thereby offering a dependable reference for making informed task offloading decisions.

\subsection{Time Delay Model}

The total time for processing tasks considers four essential components: transmission time, computation time, migration time, and backhaul time. 

\subsubsection{Satellite Computation}

We assume that the satellite operates using a single-core Central Processing Unit (CPU). Therefore, when the UE offloads computational tasks to the satellite for execution, the satellite can only execute one task at a time. Consequently, when the satellite receives multiple tasks from the UE, these tasks are computed sequentially in a first-come, first-served order.

Assuming that at time $t$, the UE offloads task $i$ to the satellite $j$ through the satellite-terrestrial transmission link, the upload time is $t_{i,j}^{upload}=D_{i}/R_{j}(t)$, where $D_{i}$ represents the task size and $R_{j}(t)$ represents the available transmission rate of the satellite $j$ at time $t$.

The computation time $t_{i,j}^{comp}$ for task $i$ on the satellite $j$ can be calculated by dividing the data size $D_i$ by the computation rate $\beta_j$ of the satellite $j$, expressed as $t_{i,j}^{comp} = D_{i} / \beta_{j}$.

\subsubsection{Local Computation and Satellite Uplink-Transmission}

When executing computational tasks on the UE, we also consider it as a single-core CPU, executing only one task at a time. This implies that multiple tasks will be computed sequentially on the UE and executed in a first-come, first-served order. For a local task $i$ with a data size $D_i$, its local computation time is $t_{i}^{comp}=D_{i}/\alpha$, where $\alpha$ denotes the computation speed of the UE. The total computation time for all local tasks is $t_{local}^{comp}=\sum^{N}_{i=1}(1-g_{i})D_{i}/\alpha$, where $g_{i} = 0$ for local tasks.

In the time-sequential offloading scheme, the transmission of a new task is delayed until the previous task completes its transmission. The end time of the UE's offloading transmission is $t^{upload,end} 
=\sum_{i=1}^{N} \sum_{j=1}^{M} \eta_{i j}t_{i,j}^{upload} $. 

Once the UE has finished transmitting tasks to be offloaded to the satellite, it can immediately starts local computation. Therefore, the start time for the UE's local computation is $t_{local}^{comp,start}=t^{upload,end}$, and the corresponding end time is $t_{local}^{comp,end}=t_{ local}^{comp,start}+t_{local}^{comp}$.

\subsubsection{Queuing Model for Offloading and Computation}

\begin{figure}[!t]
\centering
\includegraphics[width=0.8\columnwidth]{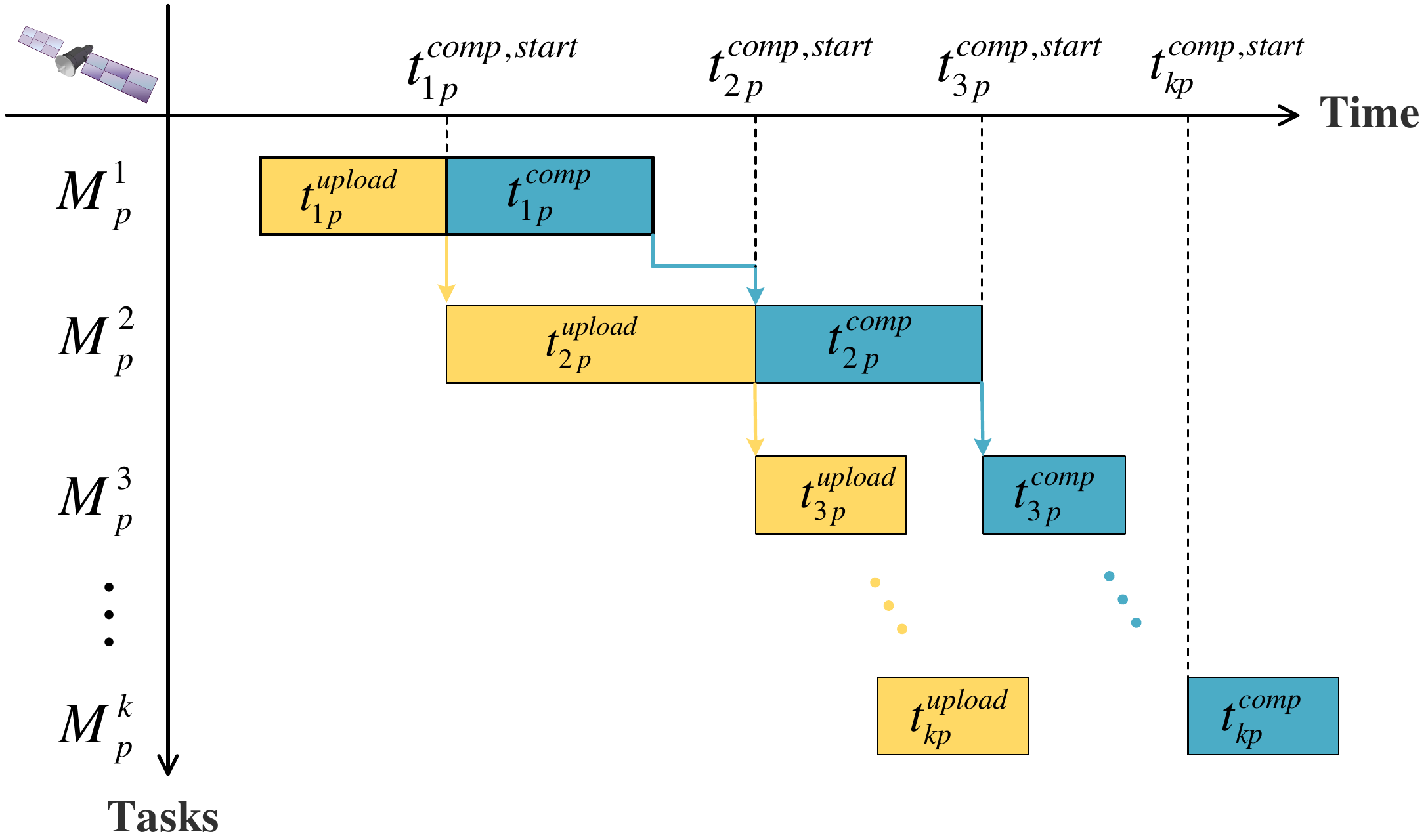}
\caption{Queuing model for task offloading and computation.}
\label{fig:Offloading Calculation Queuing Model}
\end{figure}

The offloading order of tasks plays a crucial role in the computation time. It is necessary to arrange the task offloading order rationally to minimize waiting time in the task queue. 
Let $M_{p}={M_{p}^{1}, M_{p}^{2}, \cdots, M_{p}^{k}}$ represent a set of $k$ tasks received on a satellite $p$, where $k=\sum_{i=1}^{M}\eta_{ip}$. When a task is transmitted to the satellite $p$, the satellite will start computation immediately if it is idle; otherwise, this task enters the computation queue to wait for the ongoing tasks' computation. The offloading-computation queuing model on the satellite server $p$ is illustrated in Figure \ref{fig:Offloading Calculation Queuing Model}.

For the first task $M^1_p$ received by the satellite $p$, its upload process starts at $t_{1p}^{upload,start}$ and ends at $t_{1p}^{upload,end}=t_{1p}^{upload,start}+t_{1p}^{upload}$. Since it is the first task received by the satellite $p$, there is no need for queuing to start the computation. Therefore, task $M^1_p$ starts its computation at $t_{1p}^{comp,start}=t_{1p}^{upload,end}$, and ends at $t_{1p}^{comp,end}=t_{1p}^{comp,start}+t_{1p}^{comp}$, where $t_{1p}^{comp}$ is the computation time for task $M^1_p$ on the satellite $p$.

As shown in Figure \ref{fig:Offloading Calculation Queuing Model}, for the second task $M^2_p$, task $M^1_p$ has already completed its computation by the time task $M^2_p$ completes its upload. Since the satellite $p$ is idle at this time, task $M^2_p$ can immediately start its computation. Therefore, the computation starts at the time when the upload of task $M^2_p$ ends, denoted as $t_{2p}^{comp,start}=t_{2p}^{upload,end}$.

For the third task $M^3_p$, task $M^2_p$ is still under computing when the upload of task $M^3_p$ is completed. Task $M^3_p$ needs to wait for task $M^2_p$ to completes its computation. Thus, the task $M^3_p$ starts its computation when the task $M^2_p$ completes its computation, denoted as $t_{3p}^{comp,start}=t_{2p}^{comp,end}$.

For task $M^k_p$ on the satellite $p$, once it completes uploading, the computation can immediately start if task $M^{k-1}_p$ has completed computation. In this case, the start time of computation for task $M^k_p$ is $t_{kp}^{comp,start}=t_{kp}^{upload,end}$. However, if task $M^{(k-1)_p}$ has not completed computation, task $M^k_p$ must wait until task $M^{(k-1)p}$ completes its computation. In this situation, the computation start time for task $M^k_p$ is $t_{kp}^{comp,start}=t_{(k-1)p}^{comp,end}$. Considering both cases, it is evident that the start time of computation for task $M^k_p$ depends on the latter of the end time of computation for task $M^{k-1}_p$ and the end time of upload  for task $M^k_p$, represented as $t_{kp}^{comp,start}=\max ({t_{(k-1)p}^{comp,end},t_{kp}^{upload,end}})$. Consequently, the end time of computation for task $M^k_p$ is determined as $t_{kp}^{comp,end}=t_{kp}^{comp,start}+t_{kp}^{comp}$.

\subsubsection{Migration and Backhaul Model}
Due to the high satellite mobility, the satellite might become invisible when it completes its task computation. In this situation, 
the computation result must be migrated to a satellite within the visible range through ISL. The migration process follows an order, starting from the nearest adjacent satellite and progressing towards the farthest satellite until reaching those within the visible range. The migration mechanism is designed to ensure a high success rate and stability of the backhaul transmission. Let $t_{i,j}^{comp,end}$ denote the completion time of offloading task $i$ on the satellite $j$. 
If satellite $j$ is not within the visible range at that time, the computation results need to be migrated as follows.

Firstly, the geocentric position $\gamma$ of the satellite $j$ is determined. If the geocentric position falls between $0^{\circ}$ and $180^{\circ}$, to minimize the number of migrations, the computation result will be migrated in the counterclockwise direction. Assuming that the computation result of task $i$ is migrated $\lambda _{i}$ times counterclockwise, it will be migrated to the nearest satellite $j$ within the visible range. If the geocentric position of the satellite $j$ is between $180^{\circ}$ and $360^{\circ}$, the computation result will be migrated in the clockwise direction, and the satellite where the computation result is located after migration will be $j+\lambda _{i}$.

Assuming that the data migration speed between adjacent satellites is $ V_{migrate} $, the time required to migrate the computation result of the task $i$ is $t_{i}^{\text{migrate}} = \frac{\lambda_{i}D_{i}}{10V_{\text{migrate}}}$. The migration end time of the computation result of the task $i$ is $ t_{i }^{migrate,end}=t_{i,j}^{comp,end}+t^{migrate}_{i}$. In particular, if the computation result of task $i$ does not need to be migrated, then $t_{i}^{migrate,end}=t_{i,j}^{comp,end}$.

When the migration of the task $i$'s computation result is completed, the satellite $j+\lambda_{i}$, where the computation result is located, can promptly transmit the computation result back to the UE. The time required to transmit the computation result of the task $i$ to the UE is $t_{i,j}^{download}=D_{i}/10R_{j+\lambda_{i}}(t_{i}^{migrate,end})$. Consequently, the completion of the computation result for backhaul $i$ is represented by $t_{i,j}^{download,end}=t_{i}^{migrate,end}+t_{i,j+\lambda_{i}}^{download}$. When the UE receives the computation result of the task $i$, it also indicates the completion of the task $i$. Thus, the completion time of the task $i$ is $t_{i}^{end}=t_{i,j}^{download,end}$.

Therefore, the completion time for all offloading tasks is $t^{offload,end}=\max \limits_{g_{i}=1}t_{i}^{end}$.

\subsubsection{Total Time Delay}
When the offloading transmission is completed, the UE will compute its task locally immediately. After that, the offloading tasks of satellites and the local tasks of the UE will be executed simultaneously. Therefore, the total time for all tasks to be completed is the larger of the completion time of the offloaded tasks and the completion time of the local tasks, expressed as: 

\begin{footnotesize}
%\begin{small}
\begin{equation}
T_{total}=\max (t^{offload,end},t_{local}^{comp,end}).	
\end{equation}
%\end{small}
\end{footnotesize}

\subsection{Energy Consumption Model}
The energy consumption of the UE is determined by two processes:  local computation and offloading transmission. 
For local computation, it refers to the energy consumed by the UE when executing local tasks, including the energy consumption of the computational resources inside the UE and the energy consumption of the software running to perform the computational tasks. Suppose $f$ is the CPU clock frequency of the UE, and $L$ is the number of CPU cycles to complete 1-bit computation, satisfying the relation $\alpha = f/L$. According to \cite{liu2019dynamic}, the computational power $P^{comp} = \kappa f^3 $, where $\kappa$ is a hardware-determined parameter. Therefore, the computation energy consumption is $E^{comp}=P^{comp}t^{comp}_{local}=\kappa f^{3}(\Sigma_{i=1}^{N} D_{i}(1-g_{i})) / \alpha.$

For offloading transmission, it refers to the energy consumed by the UE when offloading tasks to satellites, including the energy required to establish a connection and transmit data through wireless communication. Here, we only focus on the energy consumption of data transmission. Assuming that the transmission power is $P^{tran} $ , the transmission energy consumption is $E^{tran}=P^{tran}t^{upload}.$

Thus, the total energy consumption of the UE can be expressed as:

\begin{footnotesize}
\begin{equation}
E=E^{comp}+E^{tran}.
\end{equation}
\end{footnotesize}

\subsection{Transmission Reliability Model}
Due to the high mobility of satellites' movement, the offloading policy should exhibit fault tolerance and adaptability to the unstable wireless transmission environment. In short, it is necessary to ensure that the probability of transmission failure is within an acceptable range to ensure reliability.  Here, we do not consider the local computing tasks which are not influenced by the wireless transmission. We also neglect the backhaul transmission due to its small data size. That is to say, we focus on the offloading transmission from the UE to satellites.

As discussed in Section~\ref{sec:Satellite-Terrestrial Channel Model}, we assume that the UE offloads the task $i$ to the satellite $j$ at time $t$ with a transmission BER $b_{j}(t)$. For simplicity, we rewrite it as $b_{i} = b_{j}(t)$. Thus, the correct probability is $1-b_i$. Considering the task offloading decision, the probability of successfully transmitting 1-bit data is $ {(1-b_{i})}g_{i}$, where $g_{i} = 1$ for offloading tasks.

For each task, its successful offloading requires the correct transmission of every bit of its data. Hence, the probability of successfully transmitting the task $i$ is $r_{i}^{ {success }}=\left(\left(1-b_{i})\right) g_{i}\right)^{D_{i}}.$

For the UE, the offloading process is considered successful only if each task is transmitted successfully. Thus, its probability of successful transmission is  $r^{ {success }}=\prod_{i=1}^{N} r_{i}^{ {success }}=\prod_{i=1}^{N}(\left(\left(1-b_{i})\right) g_{i}\right)^{D_{i}}).$ If any task fails to offload, it indicates a failure in the task offloading process for the UE. Therefore, the probability of offloading failure of the UE is determined by:

\begin{footnotesize}
\begin{equation}
r^{ {failure }}=1-r^{{success }}=1-\prod_{i=1}^{N}(\left(\left(1-b_{i})\right) g_{i}\right)^{D_{i}}).
\end{equation}
\end{footnotesize}

\subsection{Privacy-Preserving Model}
During the task offloading process, we consider two crucial privacy concerns: the usage pattern privacy and the location privacy. In this paper, we focus on the privacy leakage on untrusted satellites, assuming that the UE is fully trusted and secure. 

\textbf{Usage Pattern Privacy-Preserving: }
Usually, the UE generates data of varying sizes and formats based on different usage patterns. As a result, an insecure satellite has the potential to deduce the usage pattern by analyzing the size of the data volume associated with the offloaded tasks. For instance, the satellite might extract statistical information from its task offloading history, as well as the UE's usage patterns. This information can be utilized as a unique identifier or fingerprint to detect the UE. In addition, the satellite might even pinpoint the application running at the UE when there is a certain pattern in the task properties generated by the application, which poses a significant security threat to privacy-sensitive UEs. 

To protect the UE's usage pattern privacy, the data volume can be manipulated by intentionally allowing the device to transmit redundant information. This prevents adversaries from accurately determining the actual size of the data volume in the offloading task. Assuming the current offloading decision is denoted as the $n$-th, the corresponding reward for preserving usage pattern privacy is denoted as: 

\begin{footnotesize}
\begin{equation}
\begin{aligned}
P_{u}(n)
& =  \{ x_{location}(n)>0  \}  \cdot   \{ x_{redundance}(n)=1  \}   \cdot  \{ g(n)\ge \omega   \}	\\
& +\{ x_{location}(n)=0  \}, 
\end{aligned}
\end{equation}
\end{footnotesize}
\noindent where $\left \{ \cdot \right \} $ is an indicator function, $g(n)$ means the channel state during the task offloading, and $\omega$ is a threshold value for the channel state, $x_{location}>0$ indicates the task offloading, $x_{location}=0$ indicates the local computing, and $x_{redundance}(n)=1$ indicates the intentional addition of redundant information during the task offloading. 

In the above equation, the first item means that redundant information is intentionally appended to enhance the usage pattern privacy level, when the task offloading process has a good channel state. In particular, the data size of the appended redundant information is set to 10\% of the original data size. The second item demonstrates that no redundant information will be appended for local computing.

\textbf{Location Privacy-Preserving: } 
To minimize computation delay and energy consumption, the UE tends to offload tasks to satellites with good channel conditions. However, this scheduling policy may potentially expose the UE's location privacy. The main reason is explained as follows. The channel state is highly correlated with the distances and locations among the UE and satellites. Thus, insecure satellites can infer the channel state records to obtain the UE's location. Moreover, if multiple satellites collude, they could extract an accurate location of the UE. 

To protect location privacy, the UE might intentionally offload tasks to satellites with a poor channel state to create a deceptive decision. The location privacy-preserving reward for the $n$-th offloading decision is denoted as:

\begin{footnotesize}
\begin{equation}
\begin{aligned}
P_{l}(n)
& =\left \{ x_{location}(n)>0 \right \} \cdot \left \{ g(n)<\omega  \right \} \\
& +\left \{ x_{location}(n)=0 \right \}. %\nonumber
\end{aligned}
\end{equation}
\end{footnotesize}

In the above equation, the first item means strengthening the location privacy by offloading tasks to satellites even with poor channel states, and the second item indicates no location privacy protection for local computing tasks.

\textbf{Total Privacy-Persevering: } 
In conclusion, the privacy-preserving level of the $n$-th offloading decision can be quantified as the weighted sum of the usage pattern privacy and the location privacy, denoted as $P_{total}(n)=P_{u}(n)+\varpi_{u}P_{l}(n),$
where $\varpi_{u}$ is a weighting factor reflecting the relative importance of the location privacy with respect to the usage pattern privacy. 
We calculate the total privacy-preserving level of the UE by considering the offloading decisions for all tasks, represented as: 

\begin{footnotesize}
\begin{equation}
P_{total} =\frac{1}{N} \Sigma_{n=1}^{N} P_{total}(n).
\end{equation}
\end{footnotesize}

\subsection{Problem Formulation}

Our objective is to determine an optimal policy to minimize the total offloading cost including the computation delay and the energy consumption, while satisfying pre-specified constraints on the reliability and the privacy-preserving level. It can be formulated as:

\begin{footnotesize}
\begin{equation}
\begin{aligned}
&\min\quad C=  T_{total} + \mu E_{total},\\
&\begin{array}{r@{\quad}r@{}l@{\quad}l}
s.t.&T_{total}< \hat{T},&\\
&r^{{failure }}< \hat{r},&\\
%&P_{total}\ge \hat{P}.&\\
&P_{total}\ge \hat{P},&\\
&g_{i}= \{ 0,1\}, &\forall i \in N.\\
\end{array} 
\end{aligned}
\end{equation}
\end{footnotesize}

The first constraint requires that the total completion time remains below a pre-specified threshold $\hat{T}$. The second constraint is that the total probability of the offloading failure should be less than a pre-specified reliability threshold $\hat{r}$. The third constraint is that the total privacy-preserving level must reach a pre-specified privacy level. The fourth constraint indicates that each task is either offloaded or locally computed. By satisfying these constraints, the offloading policy can effectively meet the pre-specified requirements for completion time, energy consumption, reliability, and privacy. Consequently, it can provide efficient, reliable, and privacy-preserving offloading services for user devices in real-world applications.

\textbf{Computation complexity: }
In our problem, each task should determine either local computing or remote offloading to one of the $M$ satellites. It should also determine whether to include redundant information for the privacy protection. As a result, there exist a total of $((M+1) \times 2)^{N}$ potential offloading schemes for allocating N tasks. For the time-sequential offloading scheme, the order of the transmission tasks also has an impact on the final cost. Thus, the number of offloading schemes expands further to $({((M+1) \times 2)^{N}} \times N!)$, where $N!$ represents the total permutations of $N$ tasks. Due to the huge complexity, it is infeasible to exhaustively enumerate all offloading policies to select the optimal one. Therefore, we need to develop a more efficient algorithm, which can effectively allocate tasks among the UE and multiple satellites, and determine the optimal offloading order. 

\section{Deep Reinforcement Learning based Privacy-Preserving Task Offloading}\label{sec:Privacy-preserving Task Offloading Algorithm}

In this section, we will solve the task offloading problem in the integrated satellite-terrestrial network. We formulate it as an MDP-based task assignment problem, and then propose a deep reinforcement learning approach for problem-solving. 

\subsection{Problem Description with an MDP Model}
To address the task offloading problem, we reformulate and describe it with an MDP model, where each task will be assigned to the UE or satellites according to the state space, the action space, and the reward function. Our formulation focuses on the perspective of the UE, which aims to find an optimal policy to minimize its cost function. At each time step, the UE observes the current state of the satellite-terrestrial network and selects an action based on information such as satellite positions and their workloads. Subsequently, the UE receives a reward and performs a task assignment corresponding to its chosen action. This process continues until the task assignment is completed or a constraint is violated.

In the MDP model, we denote the time step as $t$, the action performed by the UE at time step $t$ as $a_t$, the state observed by the agent at time step $t$ as $s_t$, the reward received by the agent at time step $t$ as $r_t$, and the new state entered by the agent after executing action $a_t$ from state $s_t$ as $s_{t+1}$. The fundamental module of the MDP is defined as follows:

\textbf{State space:} The state space is used to represent the environment with which the UE is attempting to interact. In our problem, the UE needs to observe the system state at each decision moment to optimally decide on the part to offload to the satellites. Consequently, the state space consists of three components: the offloading situation of all tasks, the current time, and the load situation of all satellites. We further define it as  $s_t=\{S_{task}(t),t,S_{LEO}(t)\},$
where $S_{task}(t)$ is the offloading situation of all tasks, and $S_{LEO}(t)$ is the load situation of all LEO satellites.
		
\textbf{Action space:} In our integrated satellite-terrestrial network, three decision elements need to be considered comprehensively for the behavioral decision of the UE when offloading tasks. These elements are the transmission order of tasks, the corresponding offloading location, and whether to attach redundant information. The decision action vector of UE can be defined as $a_t=\left \{  x_{num}(t),x_{location}(t),x_{redundance}(t) \right \},$
In this action vector, the first term is the task number of the current decision. The second term is the selected locations for offloading, where `0' indicates local computation and the other numbers indicate the number of the selected satellite for offloading. The third term indicates whether redundancy information is intentionally attached to the offloading, where `1' indicates that redundancy information is attached and `0' indicates no redundancy information. 

\textbf{Reward function:} The reward function serves as a feedback signal that guides the learning process based on the state representation and the chosen action. It enables the agent to adapt its actions accordingly. The primary objective of the agent is to select the action that gives the highest reward. The reward function is usually associated with an objective function, which in our formulation is to minimize the offloading cost of the UE considering the reliability and privacy-preserving level. Therefore, the value of the reward must be negatively related to the cost. When the UE makes a decision, we define the immediate reward as $r_t=\psi - C(t),$
where $\psi$ is a custom constant, and $C(t)$ denotes the total cost spent by the UE to make a decision at the time step $t$.

\subsection{PPO-Based Privacy-Preserving Task Offloading}
In our integrated satellite-terrestrial network, the task offloading policy encompasses various factors such as the computation location for each task, the transmission order of offloading tasks, and the attached options for redundancy information.
Consequently, the decision-making process involves optimizing a multi-dimensional discrete action space, which makes it challenging to solve using traditional algorithms due to the non-convex nature of the problem. 
In the above MDP model, both the state dimension and the available action space are significantly large. 

\begin{figure}[!t]
\centering
\includegraphics[width=0.85\columnwidth]{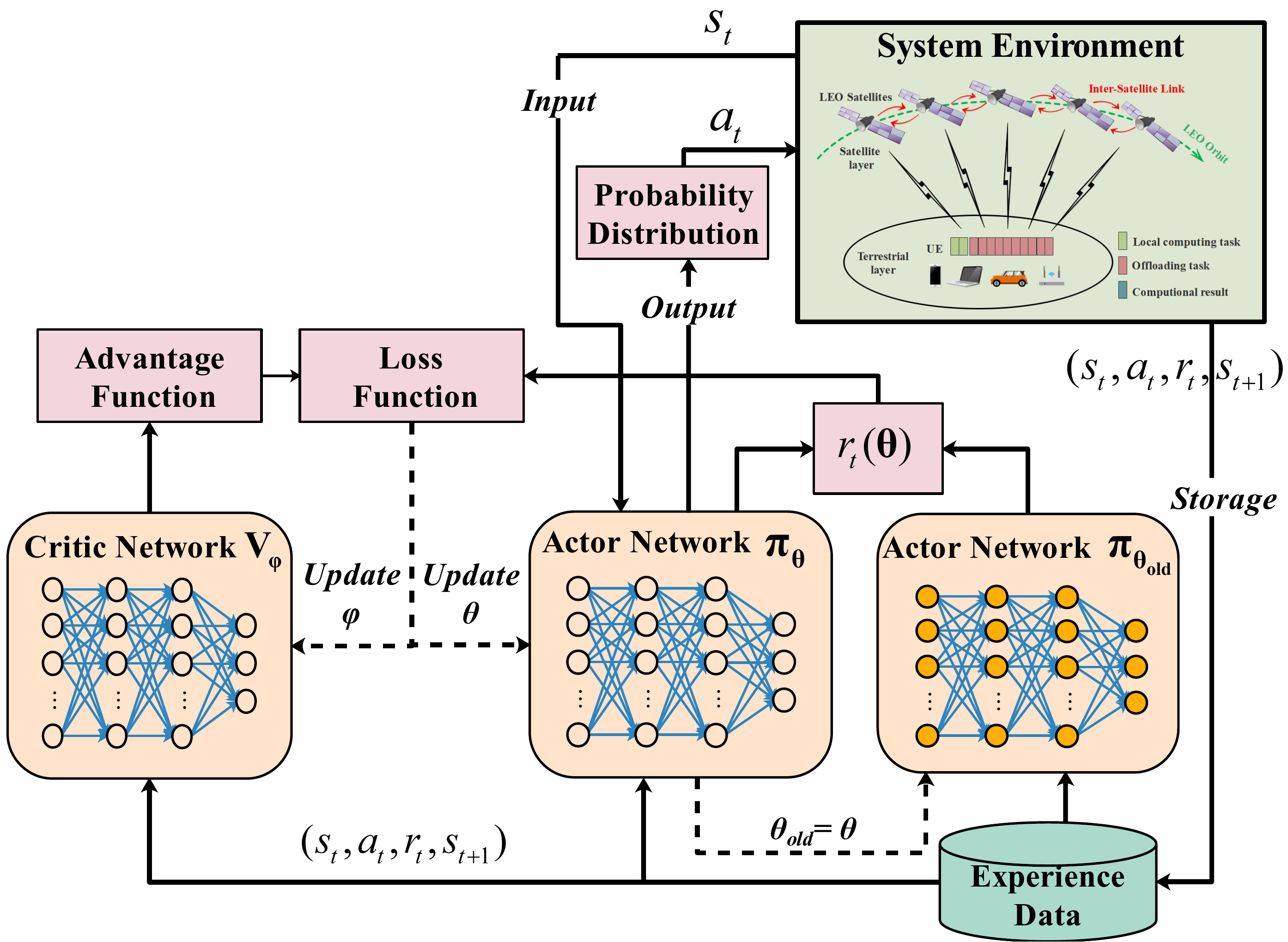}
\caption{PPO-based task offloading policy optimization.}
\label{fig:Solution principle based on PPO}
\end{figure}

Thus, we employ a deep reinforcement learning algorithm to efficiently solve the optimization problem. For the multi-dimensional discrete action spaces problem, the Proximal Policy Optimization (PPO) algorithm demonstrates good convergence, efficient sampling, greater scalability, and robustness, serving as a suitable choice to optimize our offloading policy. We implement an Actor-Critic style algorithm for task offloading policy optimization based on PPO, as depicted in Figure \ref{fig:Solution principle based on PPO}.

The PPO algorithm involves three neural networks: a new action network $\boldsymbol{\pi_{\boldsymbol{\theta}}}$ parameterized by $\boldsymbol{\theta}$, an old action network $\boldsymbol{\pi_{{\boldsymbol{\theta}}_{old}}}$ parameterized by ${\boldsymbol{\theta}}_{\boldsymbol old}$, and a value network $\boldsymbol{V_{\boldsymbol{\varphi}}}$ parameterized by $\boldsymbol{\varphi}$. The new action network $\boldsymbol{\pi_{\boldsymbol{\theta}}}$ and the old action network $\boldsymbol{\pi_{{\boldsymbol{\theta}}_{old}}}$ are utilized to generate probability distributions of actions in the current state, where the old action network $\boldsymbol{\pi_{{\boldsymbol{\theta}}_{old}}}$ is also used to restrict the changes in the new policy. The value network $\boldsymbol{V_{\boldsymbol{\varphi}}}$ approximates the state value function in the current state, and its output can be interpreted as the expected cumulative discounted reward for the current state, which evaluates the performance of the new action network $\boldsymbol{\pi _{\boldsymbol{\theta}}}$. 

The training process of PPO is outlined as follows. At the beginning of each round, the environmental parameters such as satellite positions and the number of tasks are initialized. The agent then interacts with the environment through the new action network $\boldsymbol{\pi_{\boldsymbol{\theta}}}$. Specifically, at time step $t$, the PPO algorithm takes the environment state $s_t$ as an input to the new action network $\boldsymbol{\pi_{\boldsymbol{\theta}}}$, which outputs a probability distribution over various actions. Subsequently, the agent selects an action $a_t$ based on this probability distribution and applies it to the environment, resulting in a reward $r_t$ and the next state $s_{t+1}$. This process generates experience data in the form of $(s_t, a_t, r_t, s_{t+1})$. Through multiple interactions with the environment, the agent accumulates a certain amount of experience data, which is then utilized for small-batch updates of the parameters $\boldsymbol{\theta}$ and $\boldsymbol{\varphi}$. 
These updates are performed as follows.

The parameter ${\boldsymbol {\theta}}$ is updated with a gradient ascent method, given by $\boldsymbol {\theta} \leftarrow \boldsymbol {\theta} + \alpha {\nabla _{\boldsymbol {\theta}}} {L}(\boldsymbol {\theta})$, where $\alpha \in [0,1)$ represents the learning rate, and ${L}(\boldsymbol {\theta})$ is the loss function for the new action network $\boldsymbol{\pi _{\boldsymbol {\theta}}}$. Considering that even small negative changes in updating ${\boldsymbol {\theta}}$ can lead to significant policy updates, the PPO algorithm introduces a clipped surrogate objective function for training the new action network $\boldsymbol{\pi _{\boldsymbol {\theta}}}$ as $\small {L}^{CLIP}(\boldsymbol {\theta }) = E\left [{ {\min \left ({{r_t({\boldsymbol {\theta })}  {\hat A_{t}}, clip\left ({{r_t({\boldsymbol {\theta })},1 - \varepsilon,1 + \varepsilon } }\right) {\hat A_{t}}} }\right)} }\right],$
where $r_t({\boldsymbol {\theta}})$ represents the probability ratio between the new policy and the old policy, given by $r_t({\boldsymbol {\theta })} = \frac {{\boldsymbol{\pi _{\boldsymbol {\theta }} }({a_{t}}|{s_{t}})}}{{\boldsymbol{\pi _{{\boldsymbol {\theta }_{old}}}}({a_{t}}|{s_{t}})}}$. If $r_t({\boldsymbol {\theta}}) > 1$, it indicates that the action $a_t$ under state $s_t$ is more likely in the new policy compared to the old policy. If $r_t({\boldsymbol {\theta}})$ is between 0 and 1, it suggests that the action  $a_t$ in the new policy is less likely than in the old policy. The function $clip(\cdot)$ is a clipping function that restricts the probability ratio to the interval $[1-\varepsilon, 1+\varepsilon]$ to control the speed of policy updates and prevent excessive parameter updates. $\hat A_{t}$ represents the advantage function, evaluating whether the behavior of the new policy is superior to that of the old policy.

In the PPO algorithm, the advantage function is commonly used to compute the target for action policy updates. By employing the advantage function, the PPO algorithm can adjust the advantages of different actions during action policy updates, thereby guiding the action network to generate improved action policies and enhancing the stability and convergence of the algorithm. The PPO algorithm utilizes the Generalized Advantage Estimation (GAE) approach to estimate the advantage function. The advantage function at time step $t$ can be defined as follows:

\begin{footnotesize}
\begin{equation}
\begin{aligned}
&&{\hat A_{t}} &=\sum_{l=0}^{\infty}(\gamma \lambda)^{l} \delta_{t+l}^{\boldsymbol V_{\boldsymbol {\varphi }}}\\
&& &=\sum_{l=0}^{\infty}(\gamma \lambda)^{l}\left[r_{t+l}+\gamma \boldsymbol V_{\boldsymbol {\varphi }}\left(s_{t+l+1}\right)- \boldsymbol V_{\boldsymbol {\varphi }}\left(s_{t+l}\right)\right], %\nonumber
\end{aligned}
\end{equation}
\end{footnotesize}
where $\gamma$ is a discount factor, $\lambda$ is a coefficient used in GAE, and ${\boldsymbol V_{\boldsymbol {\varphi }}}({s_{t}})$ represents the state value function evaluated by the value network.
	
The parameter ${\boldsymbol {\varphi}}$ is updated using the gradient descent method as $\boldsymbol {\varphi} \leftarrow \boldsymbol {\varphi} - \alpha {\nabla _{\boldsymbol {\varphi}}} {L}(\boldsymbol {\varphi})$. Here, ${L}(\boldsymbol {\varphi})$ represents the loss function to evaluate the network $\boldsymbol{V_{\boldsymbol {\varphi}}}$, defined by ${L}(\boldsymbol {\varphi}) = \left({ {\boldsymbol V_{\boldsymbol {\varphi}}}({s_{t}}) - R_{t}}\right)^{2},$
where $R_{t}$ denotes the cumulative discounted reward. It represents the sum of all reward values obtained by the agent during its interaction with the environment, given by ${R_{t}} = \sum_{l=0}^{\infty} {{\gamma ^{l}}{r_{t + l}}}$, where ${\gamma}$ is the discount factor.

After one iteration, the parameters of the old action network ${\boldsymbol {\theta }}_{old}$ are replaced by the parameters of the new action network ${\boldsymbol {\theta }}$. Once the parameter update is completed, the PPO algorithm clears the accumulated experience data from the training process to prepare for a new iteration. This ensures that the model learns from a fresh environment in each iteration and generates new action selections based on the latest action network parameters, thereby enhancing the performance and effectiveness of the algorithm.

The pseudo-code for the training process of the PPO-based Algorithm is presented in Algorithm \ref{alg1}.

\begin{algorithm}[!t]
\caption{Training process of the PPO-based algorithm.}
\footnotesize
\label{alg1}
\begin{algorithmic}[1]
\STATE Initialize the experience replay pool $\mathcal{D}$
\STATE Initialize the actor network $\boldsymbol{\pi _{\boldsymbol {\theta }}}$ and the critic network $\boldsymbol{V_{\boldsymbol {\varphi }}}$
\STATE Randomly initialize network parameters $\boldsymbol {\theta }$ , $\boldsymbol {\varphi }$ ; ${\boldsymbol {\theta }}_{old} \gets {\boldsymbol {\theta }}$
\FOR{each episode}
\STATE Initialize environment
\STATE Actor network $\boldsymbol{\pi _{\boldsymbol {\theta }}}$ interacts with the environment for $T$ timesteps, where it collects experience data $(s_{t}, a_{t}, r_{t}, s_{t+1})$ and stores them in the experience replay pool $\mathcal{D} $
\STATE Calculate advantage estimates as follows: \par
\quad ${\hat A_{t}} =\sum_{l=0}^{\infty}(\gamma \lambda)^{l}\left[r_{t+l}+\gamma \boldsymbol{V_{\boldsymbol {\varphi }}}\left(s_{t+l+1}\right)- \boldsymbol{V_{\boldsymbol {\varphi }}}\left(s_{t+l}\right)\right]$

\FOR{$i=1$ \TO $F$}
\STATE Calculate loss function using experience data in $\mathcal{D} $ as follows: \par
\quad ${L}^{CLIP}(\boldsymbol {\theta }) = E\left [{ {\min \left ({{r_t({\boldsymbol {\theta })}  {\hat A_{t}}, clip\left ({{r_t({\boldsymbol {\theta })},1 - \varepsilon,1 + \varepsilon } }\right) {\hat A_{t}}} }\right)} }\right]$ \\
\quad $ {L}(\boldsymbol {\varphi })={ \left({ \boldsymbol{V_{\boldsymbol {\varphi }}}({s_{t}})-R_{t}}\right)^{2} }$  \\
\STATE Update $\theta$ and $\varphi$ with a gradient method w.r.t. ${L}^{CLIP}(\boldsymbol {\theta })$ and ${L}(\boldsymbol {\varphi })$
\ENDFOR
\STATE Update $\theta_{old}$ with $\theta$
\STATE Clear the experience replay pool $\mathcal{D} $
\ENDFOR
\end{algorithmic}
\end{algorithm}

\section{Performance Evaluation}\label{sec:Performance Evaluation}
In this section, we evaluate the performance of our proposed PPO-based optimization algorithm for privacy-preserving task offloading in the integrated satellite-terrestrial network. 

\subsection{Experimental Settings}
In this experiment, we conducted our research using a simulation environment and relevant programming tools. Specifically, we utilized OpenAI Gym as the simulator, Stable Baselines3 as the reinforcement learning library, and Python 3.8.13 as the programming language. OpenAI Gym is an open-source toolkit that offers a collection of environments with a standardized interface, allowing for the creation of custom environments. During the training phase, the model and network parameters were saved for subsequent testing. The scheduling agents were implemented and trained using the Stable Baselines 3 (SB3) library, which provides a robust implementation of state-of-the-art reinforcement learning algorithms based on PyTorch. This library also facilitates the training, testing, and saving of RL agents within the OpenAI Gym environment. These tools and environments enabled us to perform simulations and evaluate the performance of different algorithms for our research problem. For our experiments, we utilized a server equipped with a 2.40 GHz Intel(R) Xeon(R) Gold 6240R CPU and an NVIDIA Corporation GV100GL [Tesla V100S PCIe] graphics card with 32 GB video memory. 

We consider a semi-circular region encompassing a UE along with 25 LEO satellites that are uniformly distributed in orbits. These satellites are situated sequentially, starting from a geocentric angle of $344^{\circ}$. The UE generates a large number of computational tasks that need to be processed under pre-specified requirements. Table \ref{table:Simulation Settings} provides a summary of the essential parameters for our experiments.

\begin{table}[!t]  
\begin{center}
\centering
\caption{Experimental settings.}  
\label{table:Simulation Settings} 
\footnotesize
\begin{tabular}{p{5.8cm}p{2cm}} \hline
\textbf{Parameter description} & \textbf{Value}\\
Task number ($N$) & $[15,30,45,60,75,90]$ \\
Task size (\(D_{i}\)) & $[400,800,1000] MB$\\
 Transmission power of UE (\(P^{tran}\)) & $5 W$ \\
 Total bandwidth (\(B\)) & $800 MHz$ \\
Noise power (\(N_{0}\)) & \(10^{-7} W\) \\
Computing speed of LEO satellites (\(\beta\)) & \(45 MBps\) \\
Computing speed of UE (\(\alpha\)) & \(30 MBps\) \\
CPU clock frequency of UE (\(f\)) &$ 3.0 GHz$ \\
%Hardware factor of UE (\(k\)) &  \(0.2 W \cdot s^3/cycle^3\) \\
Hardware factor of UE (\(k\)) &  \(0.2 W s^3/cycle^3\) \\
Reference channel gain at a distance of 1m (\(\beta _{o}\)) & $- 50 dBm$ \\
Radius of the Earth (\(R\)) & $6371 km$ \\
Orbital height of the LEO satellite (\(H\)) & $780 km$ \\
Angle between adjacent LEO satellites (\(\theta\)) &$ 2^{\circ} $ \\
Weighting factor of location privacy (\(\varpi_{u}\)) &$ 1$ \\
Weighting factor of total energy consumption (\(\mu\)) &$ 1$ \\
Transmission rate between LEO satellites (\(V_{tran}\)) & $10000 MBps$ \\
Threshold of Total time for task completion (\(\hat{T}\)) &$ 200 s$ \\
Threshold of offloading failure probability (\(\hat{r}\)) & $1\%$ \\
The pre-specified level of privacy-preserving (\(\hat {P}\)) & $60\%$ \\
Threshold for evaluating channel state conditions (\(\omega\)) &	\({10}^{-6}\) \\ 
\hline
\end{tabular}   
\end{center}   
\end{table}

\subsection{Convergence Performance of the PPO-based Algorithm}

In this section, we will evaluate the converge performances of the PPO algorithm with different learning rates. We also compare it with other common DRL algorithms. 

\subsubsection{Convergence Performance Under Different Learning Rates}

\begin{figure}[!t]
\centering
\includegraphics[width=0.55\columnwidth]{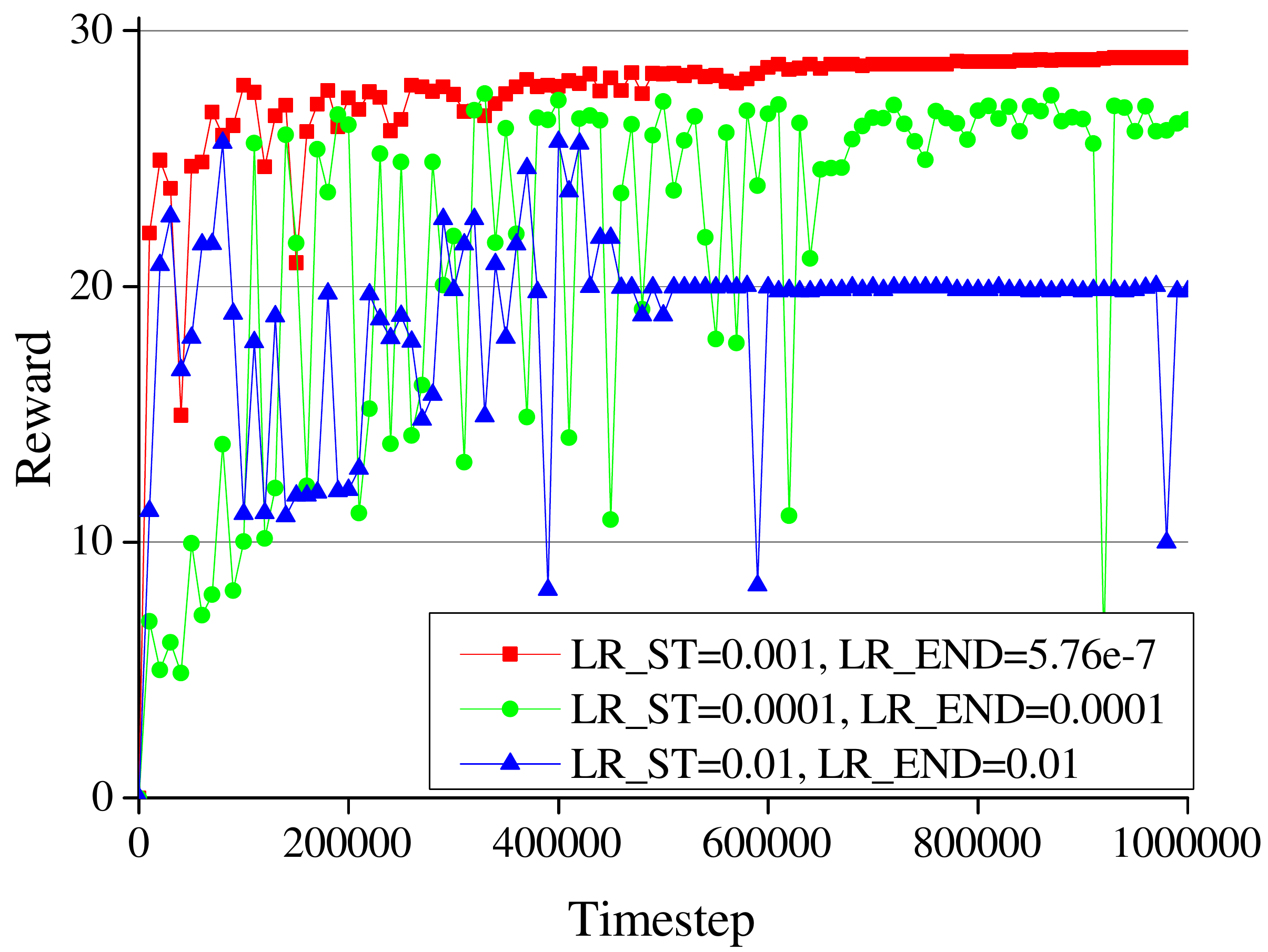}
\caption{Convergence performance under different learning rates.}
\label{fig:Result-learning rate}
\end{figure}

To investigate algorithm convergence, this paper compares the convergence curves of reward values under different learning rates. In our experiment, we set the total training time steps to $1 \times 10^6$ to effectively demonstrate the performance variations resulting from different learning rates. We employ a learning rate decay to gradually decrease the learning rate during the model training, with an initial value of 0.001 and a final value of $5.76 \times 10^{-7}$. As shown in Figure \ref{fig:Result-learning rate}, during the training time, the reward value curve gradually increases, exhibiting reduced oscillation amplitude. It converges around 700,000 steps, indicating that the agent has been trained to make an optimal decision. However, when the learning rate is fixed at 0.0001 and 0.01, the reward value curves fail to converge, and significant differences are observed compared to the learning rate decay approach. Experimental results show that the choice of learning rate significantly affects the convergence performance of the proposed PPO algorithm. By employing the learning rate decay technique, the stability, convergence, and training speed of the algorithm are improved. Therefore, in the subsequent experiments, we will adopt learning rate decay with an initial value of 0.001 to enhance the overall performance.

\subsubsection{Convergence Performance Under Different DRL Algorithms}

\begin{figure}[!t]
\centering
\includegraphics[width=0.55\columnwidth]{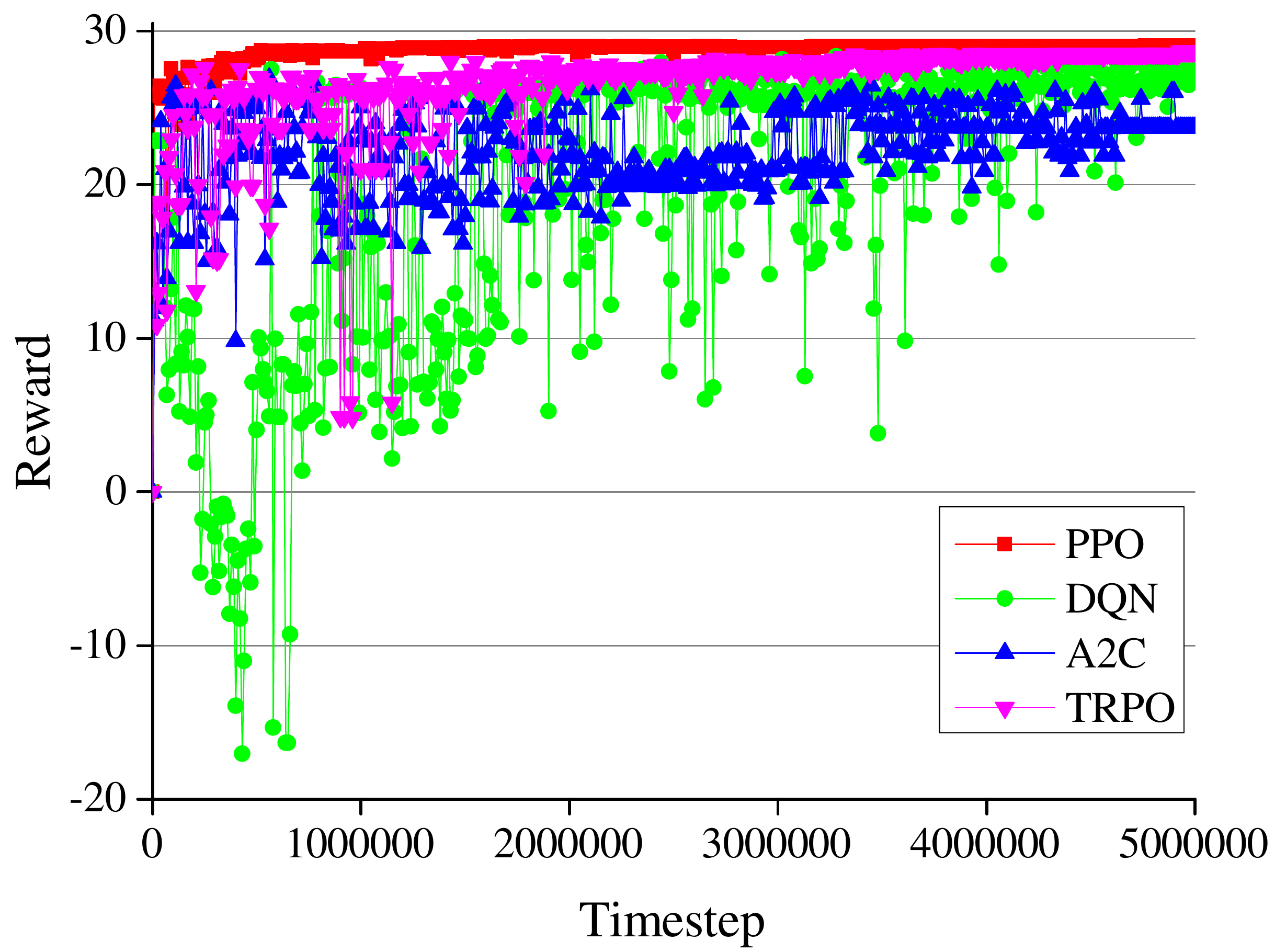}
\caption{Convergence performance under different deep reinforcement learning algorithms.}
\label{fig:Result-DRL Comparsion}
\end{figure}

We further evaluated the convergence performance of our PPO algorithm in comparison with several typical DRL algorithms including DQN, A2C, and TRPO. As depicted in Figure \ref{fig:Result-DRL Comparsion}, we observed that the PPO algorithm outperforms other algorithms, producing superior reward values after its convergence. For the convergence speed, the PPO algorithm also exhibits significantly faster and smoother convergence than the others. It can be attributed to the PPO algorithm's capability to improve data utilization through the clipping of the objective function in the policy network, which contributes to a more stable learning process. 

\subsection{Performance Analysis for Privacy-Preserving Task Offloading}
To better evaluate our proposed PPO-based task offloading algorithm, we compare it with three benchmark algorithms: DQN-based algorithm, Random Offloading algorithm, and Uniform Offloading algorithm.

\begin{figure*} [!t]
\centering
\begin{minipage}{0.3\linewidth}
\centering
\includegraphics[width=0.8\columnwidth]{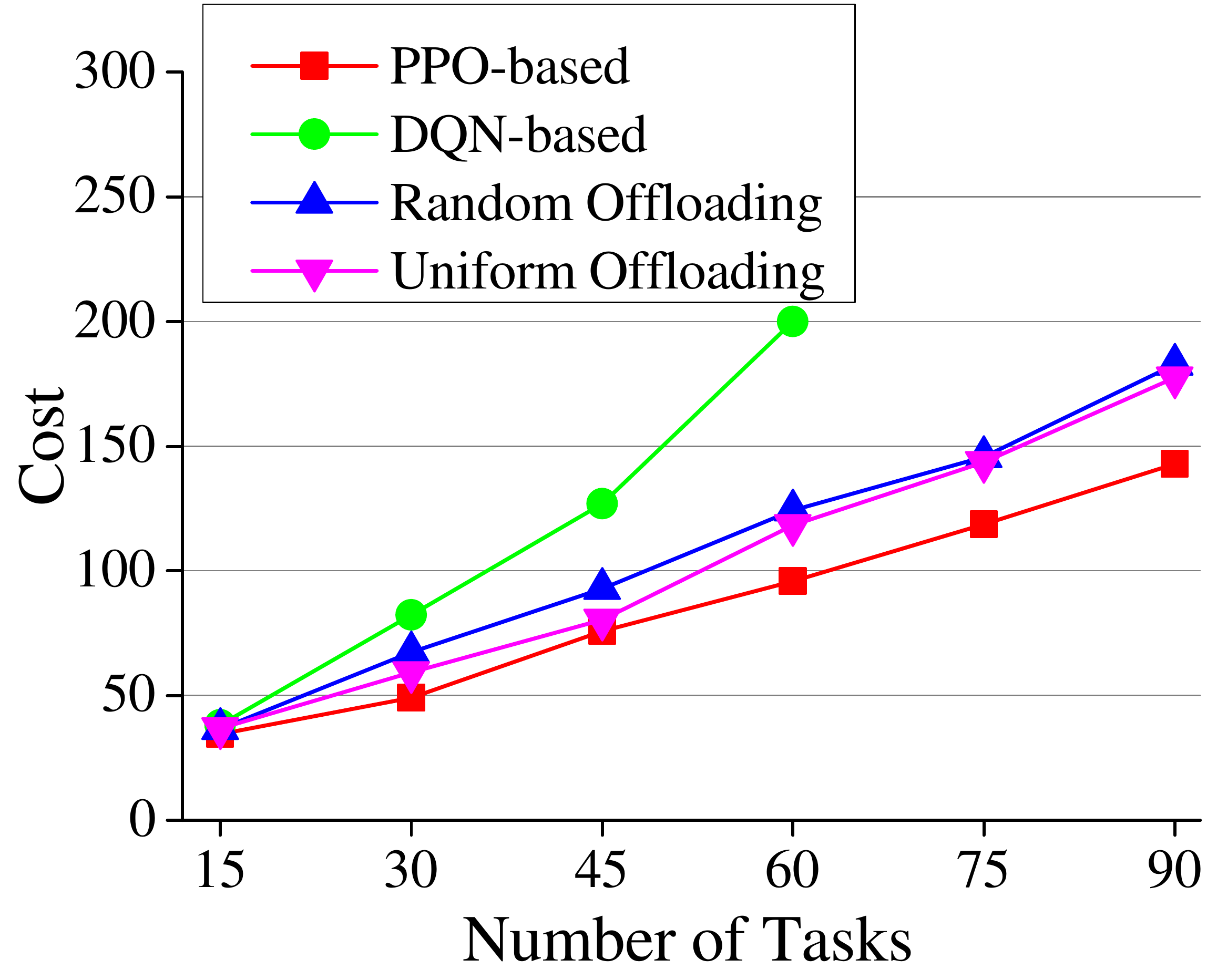}
\caption{Total cost under different numbers of tasks.}
\label{fig:Result-Number Of Task-Cost}
\end{minipage} \quad\quad
\begin{minipage}{0.3\linewidth}
\centering
\includegraphics[width=0.8\columnwidth]{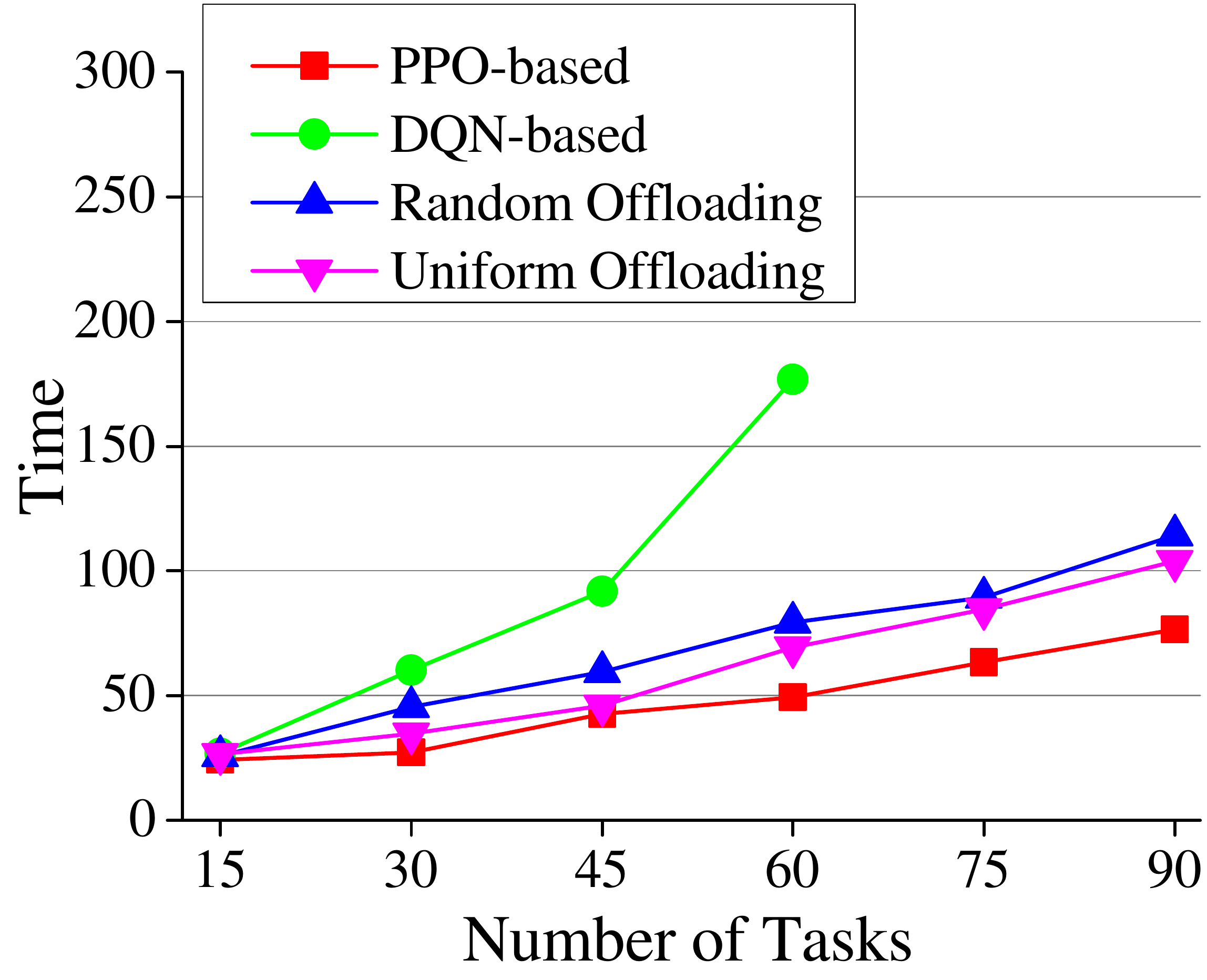}
\caption{Total time under different numbers of tasks.}
\label{fig:Result-Number Of Task-Time}
\end{minipage} \quad\quad
\begin{minipage}{0.3\linewidth}
\centering
\includegraphics[width=0.81\columnwidth]{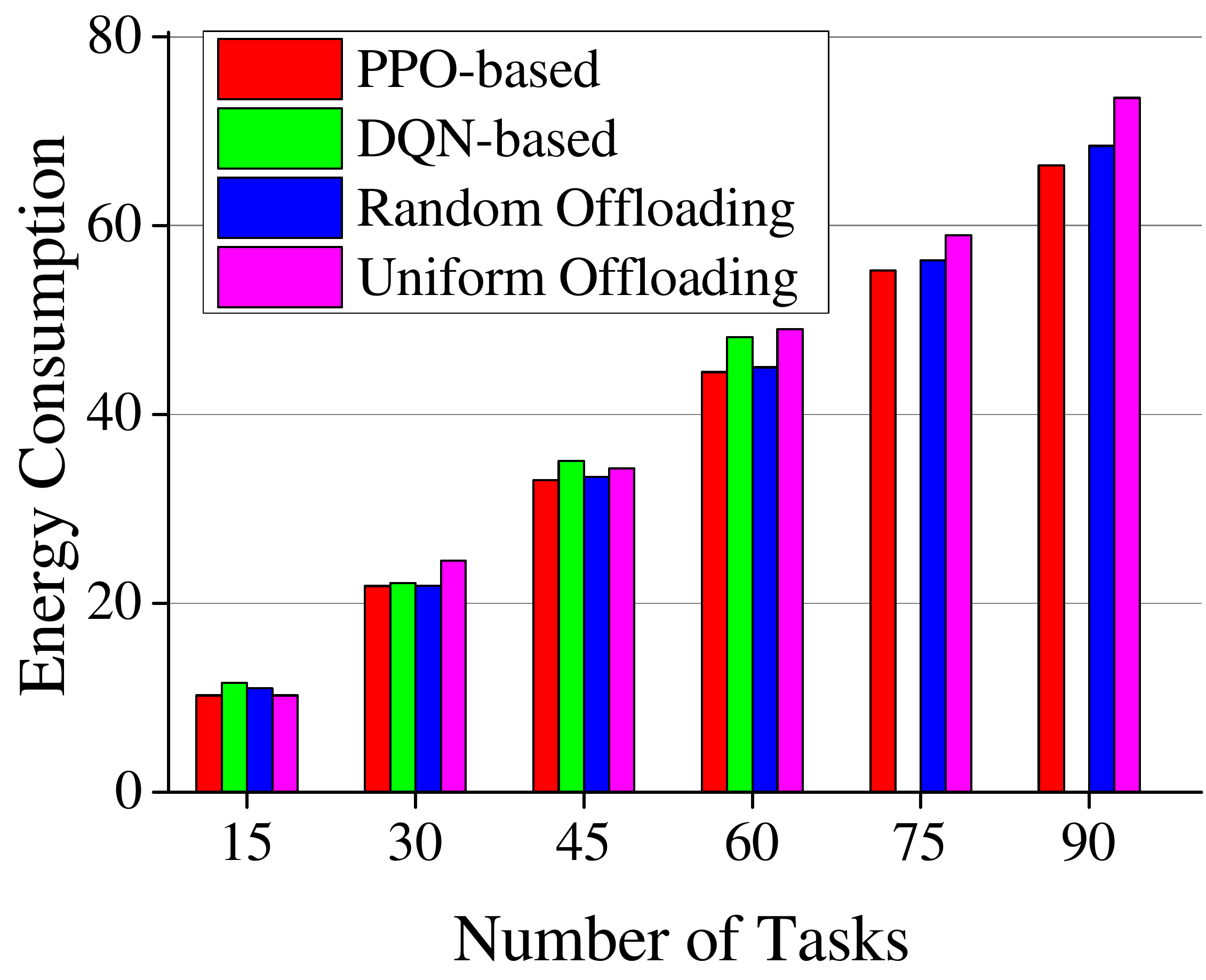}
\caption{Total energy consumption under different numbers of tasks.}
\label{fig:Result-Number Of Task-Energy}
\end{minipage} 
\end{figure*}

\textbf{DQN-based:} In this algorithm, we employ the Deep Q-Network (DQN) algorithm to optimize the task offloading problem. DQN is a value-based reinforcement learning algorithm that leverages the learning of action-value functions to determine the optimal policy. Considering the implementation of the DQN algorithm in the SB3 library, it is necessary to reduce the dimensionality of the action space in the MDP model. Specifically, the original three-dimensional action space needs to be transformed into a one-dimensional action space to satisfy the input requirements of the DQN-based algorithm and facilitate better action selection and parameter optimization during the training process.

\textbf{Random Offloading:} This algorithm generates a pool of 1,000 random offloading policies. After evaluation and comparison of these policies, the algorithm identifies and selects the policy with the best performance.

\textbf{Uniform Offloading:} 
In this algorithm, all tasks are uniformly distributed among available satellites to ensure a rough load balance for each satellite. The allocation follows a counterclockwise order on the orbital plane. 

\subsubsection{Total Cost Under Different Numbers of Tasks}
We investigate the task offloading performance with different numbers of tasks. Figure \ref{fig:Result-Number Of Task-Cost} illustrates the total cost for task numbers ranging from 15 to 90, encompassing three different task sizes (400MB, 800MB, and 1,000MB), with an equal distribution of these tasks among the total number of tasks. As shown in Figure \ref{fig:Result-Number Of Task-Cost}, as the number of tasks increases, the total costs of all algorithms increase, but the total cost of our algorithm is always the smallest. For 15 tasks, the total cost of each algorithm is almost identical. When the number of tasks is larger than 15, the differences among them become significant. With 30 tasks, our PPO-based algorithm reduces the total cost by 40.40\%, 27.20\%, and 17.15\% in comparison with the DQN-based algorithm, Random Offloading algorithm, and Uniform Offloading algorithm, respectively. It means that our proposed algorithm achieves the best decision for each task. When the task number reaches 60, the DQN-based algorithm exhibits the highest cost. This can be attributed to the fact that the DQN-based algorithm relies on experience replay and an $\epsilon$-greedy policy, striking a balance between exploration and exploitation within the action space. When the action space becomes larger, it may result in insufficient exploration or excessive exploration, thereby affecting the quality of the offloading policy. Furthermore, for a wide range of action spaces, the DQN-based algorithm may require more training data and a longer training time to discover the optimal offloading policy, thus yielding relatively inferior offloading policies and higher costs. As the task number reaches 75 or 90, the exploration becomes more challenging for the DQN-based algorithm due to the increased size of the action space, making it difficult to find offloading policies that satisfy the constraint conditions. In summary, the PPO-based algorithm shows significant performance advantages in terms of the total cost, achieving more economical and efficient task offloading policies compared to other algorithms.

\subsubsection{Total Time Under Different Numbers of Tasks}
We have conducted a study on the total time taken by different algorithms for varying numbers of tasks. Figure \ref{fig:Result-Number Of Task-Time} illustrates the total time taken by different algorithms when the number of tasks ranges from 15 to 90. It can be observed that among these algorithms, the PPO-based algorithm demonstrates superior performance. As the number of tasks increases, task transmission and computation pressures also significantly increase, requiring much more time for task transmission, computation, and migration. Consequently, the total time for all approaches also increases. However, even with an increasing number of tasks, our proposed PPO-based algorithm consistently exhibits the lowest total time. This indicates a significant advantage of our algorithm in handling a large number of tasks, effectively reducing the total time and improving system performance and efficiency.

\subsubsection{Total Energy Consumption Under Different Numbers of Tasks}
We investigate the total energy consumption of our algorithm for different numbers of tasks. The variations in total energy consumption among different algorithms are shown in Figure \ref{fig:Result-Number Of Task-Energy} where the number of tasks ranges from 15 to 90. It can be observed that the task number increment leads to more transmission and local computation, resulting in an overall increment in energy consumption for all algorithms. However, our proposed PPO-based algorithm still consistently exhibits the lowest total energy consumption, especially when the number of tasks reaches a large number (e.g., 90). It confirms the significant advantages of our proposed algorithm in handling a large number of tasks, effectively reducing the total energy consumption of the UE. 

\subsubsection{Total Cost Under Different Reliability Requirements}

\begin{figure*} [!t]
\centering
\begin{minipage}{0.48\linewidth}
\centering
\includegraphics[width=0.5\columnwidth]{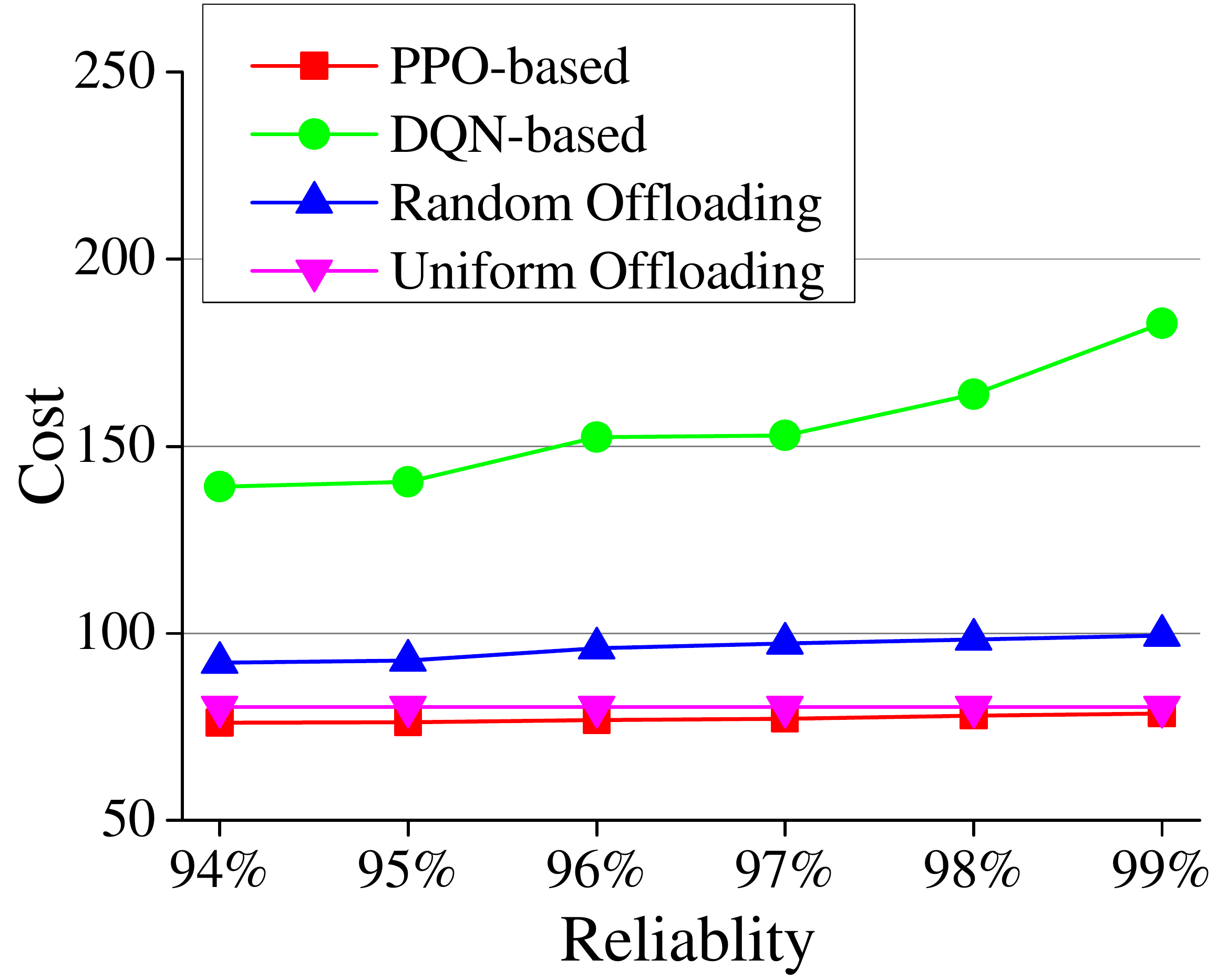}
\caption{Total cost under different reliability requirements.}
\label{fig:Result-Reliability}
\end{minipage} %\quad
\begin{minipage}{0.48\linewidth}
\centering
\includegraphics[width=0.5\columnwidth]{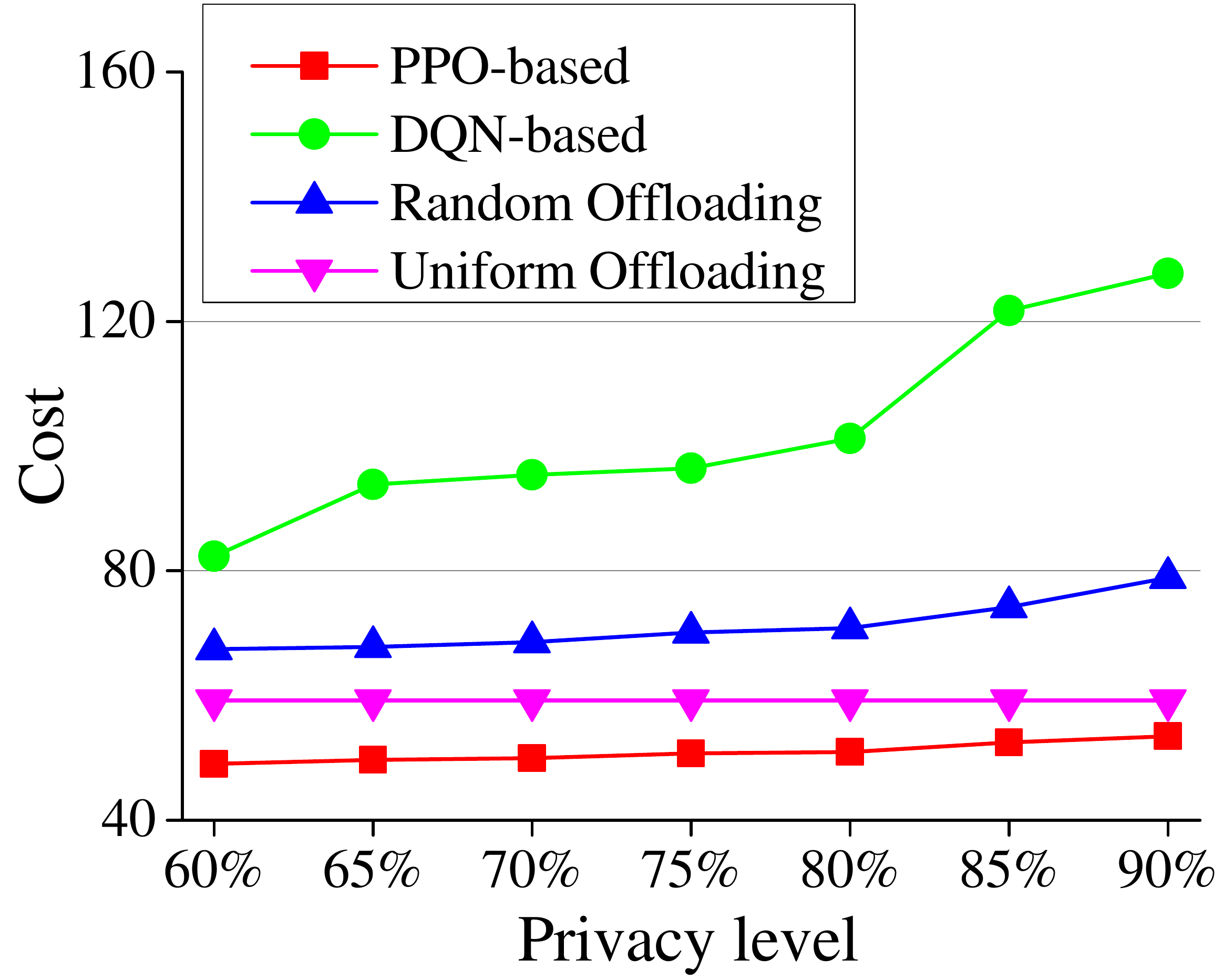}
\caption{Total cost under different privacy levels.}
\label{fig:Result-Privacy}
\end{minipage} %\quad
\end{figure*}

When the number of tasks is set to 45, the total costs for different algorithms are shown in Figure \ref{fig:Result-Reliability} with the pre-specified reliability requirement increasing from 94\% to 99\%. We can infer that as the reliability requirements increase, the total costs of the PPO-based algorithm, DQN-based algorithm, and Random offloading algorithm also increase due to their offloading policy adjustments. The main reason can be explained as follows. As the pre-specified reliability requirement gradually increases, the offloading policy undergoes a series of changes such as task re-ordering, the adjustment of offloading locations, and the addition of redundant information. Specifically, to meet higher reliability requirements, the UE may prioritize offloading tasks to satellites that are closer in proximity and have favorable channel conditions. Alternatively, it may choose to perform computations locally. These choices result in excessive workloads in the UE or certain satellites, increasing the task completion time and the total cost. Furthermore, the UE also needs to make more refined offloading decisions to ensure reliable task completion. 
It should be noted that the Uniform Offloading algorithm adopts a fixed task offloading order, so its offloading policy remains unchanged with a constant total cost, as the reliability requirements increase. However, even in this scenario, our proposed PPO-based algorithm maintains the lowest cost, demonstrating its superior performance under high reliability requirements and making it an effective choice for offloading policies.

\subsubsection{Total Cost Under Different Privacy Levels}

When the privacy-preserving level increases from 60\% to 90\%, the total cost of different algorithms are illustrated in Figure \ref{fig:Result-Privacy}. 
As the pre-specified privacy-preserving level requirements gradually elevate, a series of changes occur in the offloading policy. Specifically, to meet higher privacy-preserving level demands, the UE may prioritize task offloading to satellites located at a larger distance with poorer channel conditions, or opt for its local computation. These offloading policies will prolong task completion time and increase energy consumption.
As the privacy-preserving level rises, the total costs of the PPO-based algorithm, DQN-based algorithm, and Random offloading algorithm increase due to their offloading policy adjustments. When the privacy-preserving level requirement is set at 60\%, our proposed PPO-based algorithm reduces the total cost by 40.41\%, 27.20\%, and 17.15\% in comparison with the DQN-based algorithm, the Random Offloading algorithm, and the Uniform Offloading algorithm, respectively. 
As the privacy-preserving level requirement rises, our proposed PPO-based algorithm can still achieve a reasonable balance among task completion time, energy consumption, privacy-preserving level, and communication reliability, thereby maintaining the lowest total cost. It demonstrates the effectiveness of our PPO-based algorithm in achieving optimal offloading policies even under high privacy-preserving level requirements.

\section{Conclusion}\label{sec:conclusion}
In this paper, we seek the task offloading opportunity in a satellite-terrestrial network. Our initial target is to optimize the offloading cost, communication reliability, and the user privacy leakage in satellite-assisted edge computing scenarios, subject to certain constraints such as satellite mobility and coverage time. We formulate this problem with an MDP model, and propose a PPO-based deep reinforcement learning algorithm to achieve an optimal task offloading policy. Extensive experimental results showcase the superiority of our proposed algorithm compared to other benchmark algorithms, producing an excellent balance among task completion time, energy consumption, privacy-persevering level, and communication reliability. We believe it is an effective solution for privacy-preserving task offloading in satellite-assisted edge computing, thereby contributing to the enhancement of the quality and security of satellite-assisted edge computing services.

\section*{Acknowledgments}	
This work is supported by National Natural Science Foundation of China (No. 61802383), Research Project of Pazhou Lab for Excellent Young Scholars (No. PZL2021KF0024), and Guangzhou Basic and Applied Basic Research Foundation (No. 202201010330, 202201020162).

\begin{IEEEbiographynophoto}{Wenjun Lan}
received his Bachelor degree in computer science and technology from Wuyi University in 2021. He is currently pursuing his Master degree at Guangzhou University, China. His research interests include machine learning and mobile edge computing.
\end{IEEEbiographynophoto}

\begin{IEEEbiographynophoto}{Kongyang Chen}
is an Associate Professor at Guangzhou University, China. He received his PhD degree in computer science from the University of Chinese Academy of Sciences, China. His research interests are artificial intelligence, edge computing, blockchain, IoT, etc.
\end{IEEEbiographynophoto}

\begin{IEEEbiographynophoto}{Yikai Li}
is a master student at Guangzhou University. His research interests are mobile edge computing and federated learning.
\end{IEEEbiographynophoto}

\begin{IEEEbiographynophoto}{Jiannong~Cao}
is currently the Otto Poon Charitable Foundation Professor in Data Science and the Chair Professor of Distributed and Mobile Computing in the Department of Computing at The Hong Kong Polytechnic University (PolyU), Hong Kong. He is also the Dean of Graduate School, the director of Research Institute for Artificial Intelligence of Things (RIAIoT) in PolyU, the director of the Internet and Mobile Computing Lab (IMCL). 
His research interests include distributed systems and blockchain, wireless sensing and networking, big data and machine learning, and mobile cloud and edge computing. He has served the Chair of the Technical Committee on Distributed Computing of IEEE Computer Society 2012-2014, a member of IEEE Fellows Evaluation Committee of the Computer Society and the Reliability Society, a member of IEEE Computer Society Education Awards Selection Committee, a member of IEEE Communications Society Awards Committee, and a member of Steering Committee of IEEE Transactions on Mobile Computing. He has also served as chairs and members of organizing and technical committees of many international conferences, including IEEE INFOCOM, IEEE PERCOM, IEEE IoTDI, IEEE ICPADS, IEEE CLOUDCOM, SRDS and OPODIS, and as associate editor and member of the editorial boards of many international journals, including IEEE TC, IEEE TPDS, IEEE TBD, IEEE IoT Journal, ACM ToSN, ACM TIST, ACM TCPS.
He is a member of Academia Europaea, a fellow of the Hong Kong Academy of Engineering Science, a fellow of IEEE, a fellow of China Computer Federation (CCF) and an ACM distinguished member.
\end{IEEEbiographynophoto}

\begin{IEEEbiographynophoto}{Yuvraj Sahni}
received the PhD degree from The Hong Kong Polytechnic University, Hong Kong, in 2021. He is currently a Research Assistant Professor at The Hong Kong Polytechnic University, Hong Kong. His research interests include edge computing, IoT, and smart buildings.
\end{IEEEbiographynophoto}

\end{document}